\documentclass[prd,reprint,aps,superscriptaddress,nofootinbib]{revtex4-2}
\usepackage{graphicx} % Required for inserting images
\usepackage{amsmath,amssymb}
\usepackage{slashed}
\usepackage{cancel}
\usepackage{color}
\usepackage{hyperref}
\usepackage[english]{babel}

\makeatletter
\let\selectlanguage\@gobble
\makeatother

\makeatletter

\newcommand{\onecolumnfootnotes}{%
  \let\set@footnotewidth\set@footnotewidth@one
  \let\compose@footnotes\compose@footnotes@one
  \onecolumn@grid@setup
  \global\count\footins=1000\relax
}

\newcommand{\twocolumnfootnotes}{%
  \let\set@footnotewidth\set@footnotewidth@two
  \let\compose@footnotes\compose@footnotes@two
  \twocolumn@grid@setup
  \global\count\footins=1000\relax
}

\makeatother

\begin{document}

%TC:ignore
\title{Tremaine–Gunn Control: \\ Evading Bounds on Light Fermion Dark Matter}

\author{Joel Barir}
\email{joelbari@mail.tau.ac.il}
\affiliation{School of Physics and Astronomy, Tel-Aviv University, Tel-Aviv 69978, Israel}

\author{Diego Blas}
\email{dblas@ifae.es}
\affiliation{Institut de Física d’Altes Energies (IFAE), The Barcelona Institute of Science and Technology, Campus UAB, 08193 Bellaterra (Barcelona), Spain}
\affiliation{Institució Catalana de Recerca i Estudis Avançats (ICREA), Passeig Lluís Companys 23, 08010 Barcelona, Spain}

\author{Anubhav Mathur}
\email{amathur@tau.ac.il, a.mathur@nyu.edu}
\affiliation{School of Physics and Astronomy, Tel-Aviv University, Tel-Aviv 69978, Israel}
\affiliation{Center for Cosmology and Particle Physics, Department of Physics, New York University, New York, NY 10003, USA}

\author{Tomer Volansky}
\email{tomerv@tauex.tau.ac.il}
\affiliation{School of Physics and Astronomy, Tel-Aviv University, Tel-Aviv 69978, Israel}

\begin{abstract}
    Pauli exclusion is often regarded as imposing a model-independent lower bound on the mass of fermionic dark matter in galaxies. We show that dark-sector interactions can substantially weaken this conclusion. In particular, eV-scale fermions can form bound structures with the characteristic sizes and densities of dwarf galaxies, thereby circumventing the conventional Tremaine–Gunn mass constraint. Within these objects, the degeneracy pressure is balanced by an attractive finite-range scalar Yukawa force that is significantly stronger than gravity. The interaction acts on dwarf-galaxy scales while remaining screened on larger scales.
    We derive the evolution of the resulting dark-matter fluid and identify a viable cosmological history in which structures first form gravitationally, with the new interaction becoming active only at late times. As a proof of concept, we construct a model in which a single late-time phase transition generates the required dynamics. These results provide a new route to exceptionally light fermionic dark matter and demonstrate that dark-sector interactions can qualitatively alter conventional phase-space limits.
\end{abstract}
%TC:endignore

\maketitle

\section{Introduction}
Positive evidence for dark matter (DM) on astrophysical and cosmological length scales constitutes one of the most urgent motivations to extend the Standard Model (SM) of particle physics. As the dominant component of the matter density in galaxies, its particle nature may be revealed through the gravitational influence it exerts (for example, on the motion of stars, or through lensing~\cite{walker_dark_2013,clowe_direct_2006,de_martino_dark_2020}). For fermionic DM, this interplay was first addressed by Tremaine and Gunn (TG)~\cite{tremaine_dynamical_1979,madsen_generalized_1991}, who used the maximum phase-space density of a degenerate gas to set a lower limit on the particle mass of order $m_\psi \gtrsim 100$ eV.
Fermions near this limit form stable degenerate bound states, supported against gravity by Fermi pressure~\cite{domcke_dwarf_2015,randall_cores_2017}. The resulting density profiles are cored, and may even fit observations better than the standard cold dark matter (CDM) prescription. 
Indeed, recent studies of small-scale structure (reviewed in~\cite{bullock_small-scale_2017,tulin_dark_2018}) suggest several deviations from the minimal CDM picture, motivating models such as self-interacting DM~\cite{spergel_observational_2000,loeb_cores_2011,kaplinghat_dark_2016} and fuzzy DM~\cite{hu_fuzzy_2000,schive_cosmic_2014,hui_ultralight_2017,ferreira_ultra-light_2021}, with qualitatively similar phenomenology.

Despite its reputation as a purely kinematic limit, the TG bound is in fact a statement about dynamics, which presumes that gravity is the only force binding the dark sector.  If, instead, an attractive dark force is at play,  stars would continue to trace the gravitational well while a fermionic DM would feel a substantially deeper one, permitting Fermi momenta well beyond what stellar motions imply. This raises a question that has long been considered settled for fermions: how light can dark matter be?  
Reopening the low-mass  window would not only open a largely unexplored mass range for fermionic DM, but also  motivate detection strategies tailored to light fermions, whose particle-like signatures differ qualitatively from the coherent-field and absorption signals commonly targeted in searches for light bosonic DM~\cite{antypas_new_2022}.

Previous attempts to evade the bound have invoked a strongly coupled sector that undergoes Cooper pairing~\cite{alexander_gravitationally_2017,alexander_strongly-interacting_2021,garani_condensed_2022}, a very large number of species~\cite{davoudiasl_ultralight_2021}, a time-dependent mass~\cite{boubekeur_tremaine-gunn_2023},  or a hybrid sector where scalars dominate within dense structures~\cite{cline_neutrinos_2026}.
In this Letter, we instead introduce an attractive Yukawa force with a finite but astrophysically large range, $\mathcal{O}(\mathrm{kpc})$, of the type widely studied in related contexts~\cite{frieman_dark_1991,gubser_cosmology_2004,nusser_structure_2005,graham_cosmological_2025}.
The force is stronger than gravity, as naturally arises in many ultraviolet completions, and thereby influences structure formation on small scales. 
Although this attractive interaction induces Cooper pairing at sufficiently low temperatures, the pairing gap is exponentially suppressed relative to the Fermi energy~\cite{gorkov_contribution_1961,kapusta_neutrino_2004}.
%~\cite{gorkov_contribution_1961,fetter_quantum_2003,pisarski_superfluidity_1999,kapusta_neutrino_2004}
Consequently, the DM fluid remains in the unpaired phase within finite structures, and at the temperatures relevant to our scenario (see Supplemental Material).

We show that, for eV-scale fermions, the Yukawa interaction supports stable, self-bound configurations with the characteristic properties of observed dwarf galaxies. Following their formation from cosmological initial conditions reveals two seemingly unrelated requirements: \textit{(i)}~the new force must become dynamically important only after gravity has initiated structure formation, and \textit{(ii)}~free-streaming of the light fermions must be suppressed at earlier times~\cite{carena_cosmologically_2022}. These requirements are not necessarily independent: dynamics that delay the cosmological role of the Yukawa attraction can naturally provide the interactions needed to suppress early free-streaming. To illustrate this connection, we present a simple realization in which a single auxiliary sector accounts for both effects.

In this Letter  we focus on dwarf galaxies, which exhibit the largest inferred coarse-grained phase-space densities among DM-dominated systems and therefore set the strongest TG bounds~\cite{dalcanton_halo_2001,boyarsky_lower_2009,di_paolo_phase-space_2018,alvey_new_2020}. 
In larger halos, the Yukawa interaction is too short-ranged to bind the system as a whole, leaving gravity to govern its global dynamics. The resulting density profiles and mass–radius relations across a broader range of galactic scales will be confronted with observations in an upcoming publication.

\section{Yukawa-Bound Degenerate Cores}

For fixed halo mass and radius, lowering the fermion mass rapidly increases the required phase-space occupancy, driving sufficiently light fermionic DM into the degenerate regime. 
The Tremaine–Gunn bound can then be understood heuristically by comparing the halo escape velocity with the Fermi velocity of its constituents. For a halo of mass $M$ and radius $R$, these scale as $v_{\rm esc}\sim\sqrt{GM/R}$ and $v_F\sim(M/m_\psi^4R^3)^{1/3}$. Stability requires $v_{\rm esc}\gtrsim v_F$, yielding a lower bound $m_\psi \gtrsim 100$ eV.
An attractive force stronger than gravity increases the escape velocity, effectively replacing $G$ by the larger coupling $\alpha/m_\psi^2$, and thereby weakens the bound.

We take the DM to be Dirac fermions $\psi$ coupled to an ultralight real scalar $\phi$,
\begin{equation}\label{eq:lagrangian-ir}
    \mathcal{L}\supset - m_\psi \bar{\psi}\psi -\frac{1}{2}m_{\phi}^{2}\phi^{2}-g\phi\bar{\psi}\psi,
\end{equation}
where $g$ is the Yukawa coupling and we define $\alpha\equiv g^2/4\pi$.
As studied previously for neutrinos~\cite{stephenson_jr_neutrino_2012,smirnov_neutrino_2022,kaplan_probing_2025}, the bound states follow from the static field equation 
\begin{equation}
\label{eq:phi}
    (\nabla^2 - m_\phi^2)\phi = g \left< \bar{\psi} \psi\right>\,,
\end{equation} 
in which $\phi$ is sourced by the fermions and shifts their mass to $m_{\psi,\text{eff}} \equiv  m_\psi + g\phi \leq m_\psi$. In the regime of interest this shift is small, $|g\phi| \ll m_\psi$, so that $m_{\psi,\text{eff}} \approx m_\psi$, and the fermions are everywhere non-relativistic, so that $\left< \bar{\psi} \psi\right>$ reduces to the number density $n_\psi$. The large mode occupation at the relevant densities further lets us treat $\phi$ as a classical field, sourced by $n_\psi$. We set the mediator range by choosing $m_\phi \sim \text{kpc}^{-1} \approx 6\times10^{-27}\text{ eV}$, so that the force dominates within dwarf galaxies, where the TG bound is ordinarily set, but is screened on the larger scales of the Milky Way and clusters.
The core then follows from the closed set of equations described by the field equation above, together with the requirement for hydrostatic equilibrium, 
\begin{equation}
\label{eq:hydro}
    \frac{dP_\psi}{dr} = - g n_\psi\left(\frac{d\phi}{dr}\right)\,,
\end{equation} 
where $P_\psi$ is the pressure of a degenerate Fermi gas (see Eq.~\eqref{eq:P} in the Supplemental Material).
Gravitational and relativistic corrections to this equation are also included in the Supplemental Material, and are found to be subdominant within the range $m_\phi^{-1}$ across our parameter space.

\begin{figure}
    \centering
    \includegraphics[width=\columnwidth]{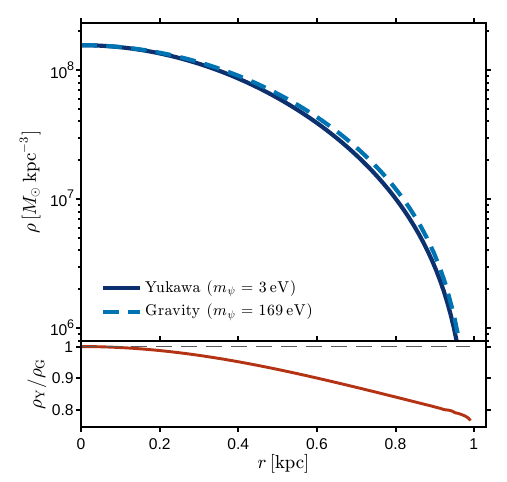}
    \caption{
    A numerical DM density profile of a dwarf-galaxy-like equilibrium configuration, obtained by solving Eqs.~\eqref{eq:phi} and~\eqref{eq:hydro}, is shown with a {\bf solid dark blue} line. The Yukawa-supported configuration has total mass $M=10^8 M_\odot$ and radius $R=1\,\mathrm{kpc}$, for the benchmark parameters $m_\phi^{-1}=1\,\mathrm{kpc}$, $\alpha=3.7\times10^{-51}$, and $m_\psi=3\,\mathrm{eV}$. Its central region forms a degenerate fermionic core supported by Fermi pressure against the attractive Yukawa interaction. For comparison, the {\bf dashed light blue} curve shows the gravity-only profile, following Ref.~\cite{domcke_dwarf_2015}, with the same central density and radius. This configuration requires $m_\psi=169\,\mathrm{eV}$ and total mass $M=5.5\times10^7 M_\odot$. The lower panel shows the ratio of the Yukawa-supported to gravity-only density profiles.
    }
    \label{fig:profile}
\end{figure}

We focus on solutions for a degenerate core at zero temperature, and have chosen the force range to ensure that dwarf galaxies are objects of this kind\footnote{This is done for consistency with the gravitational case \cite{domcke_dwarf_2015,randall_cores_2017}. Virialization may endow these structures with a finite temperature approaching the Fermi energy, but that would affect the gravity- and Yukawa-dominated profiles alike.}.  
The non-relativistic core obeys a Chandrasekhar-like mass--radius relation, obtained from the familiar gravitational result $M R^3 \sim G^{-3} m_\psi^{-8}$ by the substitution $G \to \alpha/m_\psi^2$ introduced above, yielding $M_c \sim (\alpha^3 m_\psi^2 R_c^3)^{-1}$ provided that the core is no greater than $m_\phi^{-1}$.
The dependence on radius is thus inherited directly from the gravitational case, and is similarly in conflict with the empirical mass--radius relation $M_c \propto R_c^2$ observed for galactic halos across a wide range of sizes~\cite{burkert_structure_2015,kormendy_scaling_2016,sanchez_almeida_implications_2025}.
We stress, however, that the true structure of the late-time objects will depend substantially on their entire formation history (and on gravity, within larger systems where the Yukawa force may be partly screened). 

One possibility, seen in related treatments of purely self-gravitating systems, is a virialized state (potentially of quasi-particles) consisting of a core surrounded by a dilute, approximately isothermal halo~\cite{chavanis_statistical_2020,chavanis_predictive_2022,arguelles_novel_2018,arguelles_novel_2019,arguelles_formation_2021}. 
In this picture, Milky Way sized galaxies would host a small Yukawa-bound core at their center even as their mass is dominated by the gravitationally bound halo.
These effects may modify the predicted mass--radius relation,  as would any additional interactions present in the dark sector. We therefore limit our scope to dwarf galaxies, and defer a detailed treatment of larger structures to an upcoming publication.

A dwarf-galaxy profile for our benchmark $m_\psi=3\text{ eV}$ is shown in Fig.~\ref{fig:profile}.
A profile for the purely gravitational case of Ref.~\cite{domcke_dwarf_2015} is presented for comparison; it is clear that the shape is quite similar in both cases. 

\section{Structure formation}
A complete account of how these degenerate states condense out of a homogeneous background requires an $N$-body treatment, which we leave to future work. The essential features, however, follow from a few simple estimates.  Here, we are agnostic to the production mechanism for the DM, as well as its history before equality, and require only that $\psi$  is its dominant component throughout matter domination.

Because the Yukawa force is much stronger than gravity, the free-fall time of the fermion fluid, $\tau_{\text{ff}} \sim 1/\sqrt{(\alpha/m_\psi^2) \rho_\text{DM}}$, is far shorter than a Hubble time during matter domination, so collapse occurs rapidly once the force is active. Its scale dependence follows from the linear growth of a density mode $\delta_k$ described by~\cite{nusser_structure_2005,domenech_halo_2023}
\begin{equation}
    \ddot{\delta}_k + 2H\dot{\delta}_k + \omega^2\delta_k = 0 \,,
\end{equation}
where
\begin{equation}
    \label{eq:omegasq}
    \omega^2(k) = c_s^2 \left(\frac{k}{a}\right)^2 -4\pi G \overline{\rho} \left(1 + \frac{\alpha}{G m_{\psi,\text{eff}}^2} \frac{(k/a)^2}{(k/a)^2+m_\phi^2} \right)\,,
\end{equation}
with $c_s$ and $\bar{\rho}$ the sound speed and mean mass density of the fluid respectively, and scale factor $a=1/(1+z)$. The growth of a mode is governed by $\omega^2$, which balances the pressure $c_s^2(k/a)^2$ against the destabilizing attraction. Unlike in CDM, the former term is significant since $c_s = \bar{v}_F/\sqrt{3}$ for a degenerate gas (with $\bar{v}_F \sim (\bar{\rho}/m_{\psi,\text{eff}}^4)^{1/3}$ the Fermi velocity of the homogeneous fluid). The latter is enhanced by the Yukawa contribution (the last term of \eqref{eq:omegasq}), whose reach is set by the force range. For length scales larger than the range, $k/a \ll m_\phi$, this enhancement vanishes and $|\omega^2| \lesssim 4\pi G \bar{\rho} \sim H^2$ such that the mode grows at the same slow rate as in $\Lambda$CDM (and retaining the pressure term would only slow it further).
For scales at or below the range, $k/a \gtrsim m_\phi$, the enhancement saturates to its full strength $\alpha/m_\psi^2\gg G$, so $|\omega^2| \gg H^2$ and growth proceeds on the much shorter free-fall time. Pressure arrests this growth only  up to the Jeans wavelength $\ell_J$ which, by construction, is just below the range; collapse is therefore rapid for modes between $\ell_J$ and $m_\phi^{-1}$. 
Each such region is expected to virialize through the collisionless violent relaxation of Lynden-Bell~\cite{lynden-bell_statistical_1967,chavanis_degenerate_1998}, settling into the degenerate equilibrium described above.
Its ultimate mass $M\sim \overline{\rho}(z_\text{form})\times (m_\phi^{-1})^3$ is determined by the amount of DM enclosed at the redshift of collapse, $z_\text{form}$.

For this picture to hold, there are two stipulations on the cosmological history:

\textit{(i)} Because collapse is rapid, the formation redshift $z_\text{form}$, and with it the mass $M$, is fixed by whenever the force first turns on. An onset near matter–radiation equality, in the dense early Universe, would put the bound objects at $\sim 10^{12} M_\odot$, far too massive to be the observed dwarfs.
Reproducing the desired density ($10^8 M_\odot$ within $\sim1$ kpc~\cite{walker_dark_2013,burkert_structure_2015}) instead requires the force to remain inactive until the homogeneous DM density drops sufficiently below this value that bound states attain it after virialization. This happens at $z_\text{form} \sim 100$\footnote{Baryonic accretion should not be significantly affected by this delay because it is already suppressed at earlier times by Compton drag~\cite{barkana_beginning_2001}, and because the corresponding infall timescale is still sufficiently short: $\tau_\text{infall} \sim 1/\sqrt{GM/R^3}\sim H^{-1}(z=100)$. However, the details are not studied here.}.

\textit{(ii)} Prior to $z_\text{form}$ the fermions generically constitute a non-interacting gas that free-streams at velocity $\bar{v}_F$ (which can be as large as $\mathcal{O}(1)$ at matter-radiation equality due to the substantial homogeneous density). 
As with warm DM~\cite{bode_halo_2001}, this suppresses power below distances of order $\ell_\text{fs} \sim \bar{v}_F/aH$~\cite{lesgourgues_massive_2006,carena_cosmologically_2022}. The affected scales are large enough to erase the primordial perturbations seeding dwarf galaxy-sized objects even before the Yukawa becomes active, preventing the desired structures from forming.
Circumventing this requires an interaction among the fermions that suppresses early free-streaming.
Below we ensure that the DM fermions are never free particles at the cosmological times of interest, being instead consolidated into microscopic structures which travel much more slowly than $\bar{v}_F$.
Such early-time dynamics avoid the self-interaction bounds from the Bullet Cluster~\cite{markevitch_direct_2004,randall_constraints_2008}.

Both conditions can be met by a single trigger sector which delays the onset of the force and, at earlier times, binds the fermions into slow-moving degrees of freedom that protect the seed perturbations. Its construction is outlined after a discussion of the constraints on this scenario.

\begin{figure*}
    \centering
    \includegraphics[width=1\textwidth]{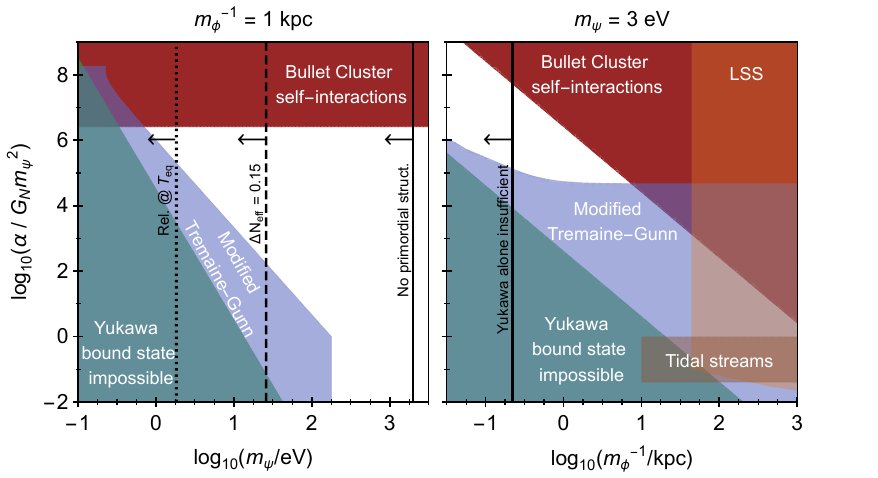}
    \caption{
    Viable parameter space for the strength of the attractive Yukawa interaction relative to gravity, with respect to the DM mass for a fixed force range $m_\phi^{-1}=1\,\mathrm{kpc}$ ({\bf left}) and with respect to the mediator range for a fixed DM mass $m_\psi = 3\,\mathrm{eV}$ ({\bf right}).
    The vertical axis is shared across both plots.
    Stable self-bound configurations are energetically allowed everywhere outside the {\bf green region}~\cite{stephenson_jr_neutrino_2012,smirnov_neutrino_2022}; within the {\bf purple region}, however, the attraction cannot sufficiently overcome Fermi pressure to reproduce the observed densities of dwarf galaxies. The {\bf red region} is excluded by constraints on coherent DM self-interactions from the Bullet Cluster~\cite{bogorad_coherent_2025}, while, well above the kpc scale,
    the {\bf orange and brown regions} mark where limits from large-scale structure~\cite{bottaro_unveiling_2024, bottaro_100_2025} and tidal streams~\cite{kesden_tidal_2006} are operative.
    The {\bf vertical line} on the {\bf right plot} indicates the force range below which only sub-kpc bound states can be formed under the Yukawa interaction alone (although gravitational accretion may still allow these smaller structures to match observed dwarf galaxies).
    On the {\bf left plot}, {\bf vertical lines} indicate representative early-Universe constraints obtained under the assumption that $\psi$ is already present as a free, homogeneous population. To the left of the {\bf dotted line}, $\psi$ remains relativistic during part of matter domination; to the left of the {\bf dashed line}, its energy density during BBN exceeds the allowed dark-radiation abundance~\cite{yeh_lbt_2026}; and to the left of the {\bf solid line}~\cite{carena_cosmologically_2022}, free-streaming erases the smallest observed cosmological structures. These early-Universe constraints are not intrinsic to the late-time bound states and can be readily relaxed through sufficiently late DM production and additional dark-sector interactions, as discussed in the text and demonstrated by the explicit model presented below.
    }
    \label{fig:parspace}
\end{figure*}

\section{Viable parameter space} 
The viable region for fermionic DM below the Tremaine–Gunn bound is shown in Fig.~\ref{fig:parspace} for the benchmarks $m_\phi^{-1} = 1\text{ kpc},m_\psi = 3\text{ eV}$. 
Three constraints on the force strength delineate this region, independently of the early cosmological history. At large coupling, the Yukawa force acts coherently over its interaction range and can displace the DM component relative to the baryons in the Bullet Cluster; the red region is excluded by requiring this displacement not to exceed the observed ${\sim}10\,\mathrm{kpc}$ separation~\cite{markevitch_direct_2004,clowe_direct_2006,bogorad_coherent_2025}. At small coupling, by contrast, the attraction is insufficient to support the desired degenerate dwarf configurations. 
The purple region contains parameters for which no bound object exists with mass $10^8\,M_\odot$ and radius at or below\footnote{A solid line marks the minimal force range for which kpc-sized objects can be formed by the Yukawa alone; below it, only smaller sized profiles are possible under this assumption.} $1\,\mathrm{kpc}$, providing the analogue of the Tremaine–Gunn bound in our model. This region is cut off at high $m_\psi$, precisely where the gravitational TG bound is satisfied (for this mass and radius) regardless of the presence of the Yukawa.
The green region marks the absolute binding threshold: for $\alpha m_\psi^2/m_\phi^2\lesssim10$, no self-bound configurations exist at all~\cite{walecka_theory_1974,stephenson_jr_neutrino_2012,smirnov_neutrino_2022}.
Note that although the new force is screened in structures much larger than dwarf galaxies, it is not known how the phase space distribution within these objects is modified since they cannot be described as degenerate cores (see the discussion concerning the profiles above). For this reason, we do not show the surviving (gravitational) TG bound arising from the Milky Way or from galaxy clusters.

The remaining constraints originate not in the late-time structures themselves but in the Fermi momentum $p_F=(3\pi^2 n_\psi)^{1/3}$ of a free fermion gas, and are therefore contingent on the early dark history. In Fig.~\ref{fig:parspace} they are distinguished by being marked with lines rather than as solid regions. 
For $\psi$ to behave as DM throughout matter domination, it must be non-relativistic by matter–radiation equality, $p_F(z_\text{eq} \approx 3400) \lesssim m_\psi$; this fails left of the dotted line. 
Further, while relativistic, the fermions contribute $p_F^4/(4\pi^2)$ to the radiation density. This exceeds the bound on $\Delta N_\text{eff}$ during nucleosynthesis~\cite{yeh_lbt_2026} left of the dashed line. 
Finally, as discussed above, the Fermi velocities will erase small-scale structure; the solid line marks where the free-streaming wavenumber $2\pi/\ell_\text{fs}$ is small enough to observably modify the power at scales below $\approx50\,h\,\text{Mpc}^{-1}$, probed by the subhalo mass function~\cite{des_collaboration_constraints_2021,carena_cosmologically_2022}\footnote{This is even more stringent than free-streaming erasing the primordial seeds of dwarf galaxies themselves, due to the radius of the corresponding comoving patches at early times.}. 
All three of these are relevant only while $\psi$ remains a free gas. If clumping into microscopic bound states renders the bulk matter cold before equality, $\psi$ will be non-relativistic at equality and can no longer free-stream, so neither the dotted nor the solid lines should apply. Conversely, delaying the production epoch to post-BBN will entirely eliminate the dashed bound on $\Delta N_\text{eff}$. With these removed, only the requirements on the force strength bound the viable region, and the figure's benchmark lies well within it.

A recent systematic treatment of strong scalar-mediated forces~\cite{graham_cosmological_2025} adds no further restriction.
Because the force is dynamically negligible before $z_{\rm form}$, the scalar has no cosmologically relevant homogeneous expectation value at recombination, leaving the cosmic microwave background unaffected. The explicit two-scalar realization presented below contains a small residual expectation value, but it is too suppressed to alter this conclusion. 
Late-time probes likewise do not reach the relevant regime\footnote{
The cosmological abundance of massive objects contributes Poisson noise to the power spectrum, which can be significant for some dark sector models, such as primordial black holes formed out of radiation~\cite{afshordi_primordial_2003}. However, if the structures are formed by local rearrangement of the homogeneous matter background, they cannot seed power on scales $\lambda$ much larger than their mean separation $d$. Indeed, due to mass and momentum conservation, it can be shown that the induced density contrast is at most $\delta \sim (\lambda/d)^2$~\cite{peebles_large-scale_1980}.}:
the Lyman-$\alpha$ forest is only sensitive to scales above $\sim0.5\,\mathrm{Mpc}$~\cite{murgia_non-cold_2017}, which exceeds the mean separation of dwarf galaxies at that redshift, while ultra-faint dwarfs constrain only clumps denser than $\sim10\,M_\odot/\mathrm{pc}^3$~\cite{graham_constraints_2024}, compared with $\sim0.1\,M_\odot/\mathrm{pc}^3$ for our Yukawa-bound structures.

\section{A trigger sector}
Requirements \textit{(i)} and \textit{(ii)} imposed by structure formation, call for a force that remains dynamically negligible in the early Universe and becomes important around $z_{\rm form}$, a behavior that can arise naturally from a low-scale phase transition.
A viable microscopic realization must, however, do more than arrange this timing: it must also stabilize the ultralight mediator mass required for a kiloparsec-range force. Quantum corrections can be naturally suppressed if $\phi$ is a pseudo-Goldstone boson of a confining sector, as familiar from QCD and composite-Higgs models~\cite{gasser_chiral_1984,panico_composite_2016,schmaltz_little_2005}. 
Finite-density effects are more subtle, however, as the ambient fermion population generically induces sizable, redshift-dependent corrections to the mediator mass that cannot be canceled at all epochs by any fixed tuning. Additional interactions introduced to control when the force becomes important will generally contribute further to these medium effects. 
Although satisfying these requirements simultaneously is nontrivial, it is certainly possible.  As a proof of concept, we present a simple construction that does so, with details relegated to the Supplemental Material.

Like other theories of ultralight scalars, our proposal is fine-tuned to impact the scales of interest. We further assume the hidden sector to be much colder than the Standard Model, so that thermal corrections do not spoil the dynamics; since the scalars are decoupled, their thermal history is unconstrained and a low temperature is readily accommodated.

We introduce a $\mathbb{Z}_2$-symmetric auxiliary scalar $\chi$, adding to the Lagrangian of Eq.~\ref{eq:lagrangian-ir} the terms
\begin{align}
    \mathcal{L} \supset -\frac{1}{2}m_\chi^2\chi^2 -\frac{1}{4}\lambda \chi^4 + \frac{1}{2F}\chi^2\bar{\psi}\psi - \frac{\kappa}{2}\phi^2\chi^2,
\end{align}
with $\lambda, \kappa\ll 1$, a bare mass $m_\chi$, and a heavy suppression scale $F\sim\text{TeV}$. The dimension-five coupling gives $\chi$ a density-dependent negative mass-squared, $\delta m_\chi^2\sim-\left< \bar{\psi}\psi \right>/F$, which dominates the bare quantity early and drives $\chi$ to a vev $\chi_*^2 \sim |\delta m_\chi^2|/\lambda$. Through the $\phi^2 \chi^2$ portal, this raises the mediator mass by $\delta m_\phi^2 \sim \kappa\chi_*^2 \sim (10^{-8}\text{ eV})^2(z/z_\text{form})^3 \gg m_\phi^2$, compressing the force's range so that it is inoperative on astrophysical scales.  As the density falls, the bare term overtakes $\delta m_\chi^2$. Choosing $m_\chi^2 \sim n_\psi(z_\text{form})/F \sim (10^{-9}\text{ eV})^2$ places this crossover around $z_\text{form}$, at which time $\chi$ returns to the symmetric phase, $\phi$ recovers its bare mass, and the force switches on, realizing \textit{(i)}.

\begin{figure*}
    \centering
    \includegraphics[width=1\textwidth]{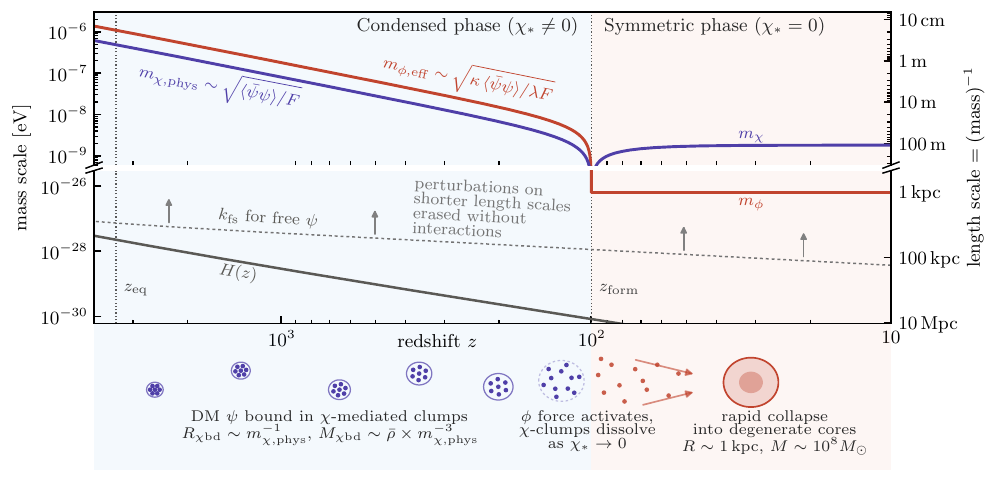}
    \caption{An illustration of the trigger sector described above. The effective ranges of the $\chi$ and $\phi$ mediated interactions are shown by the \textbf{blue} and \textbf{red} lines, respectively, for the benchmark parameters $m_\psi = 3\text{ eV},m_\phi=(1\text{ kpc})^{-1},z_\text{form}=100,F=1\text{ TeV}, \kappa/\lambda=10$. At early times, the attraction mediated by $\chi$ is enhanced due to its nonzero expectation value, which also ensures that the $\phi$ force range is reduced during this epoch. After $z_\text{form}$, the background field returns to zero, allowing $\phi$ to take on an astrophysical range while making $\chi$ irrelevant to the subsequent dynamics. Throughout the evolution, compositeness in the $\psi$ fluid preserves primordial perturbations even below the free-streaming scale in the non-interacting limit (\textbf{dashed}). 
    }
    \label{fig:uv-illustration}
\end{figure*}

The operator  $\chi^2\bar{\psi}\psi$ also supplies \textit{(ii)}, suppressing the DM free-streaming length.  In the condensed phase, it reduces to a Yukawa interaction of strength $\sim \chi_*/F$ between fermions, present only at early times. 
The range of this force, set by the in-medium mass $\lambda \chi_*^2$, never exceeds $ m_\chi^{-1} \sim 0.1\text{ km}$ while active. Though much shorter than astrophysical length scales, this encloses many particles, which the attraction rapidly binds into composite states. In this way, the matter can move far more slowly than $\bar{v}_F$, such that the dwarf-scale seeds survive. 
Since both the strength and range of this interaction are set by $\chi_*$, it switches off continuously as the density falls, and these early clumps disintegrate precisely when the long-ranged Yukawa becomes active.
See Fig.~\ref{fig:uv-illustration} for an illustration of the dynamics, and the Supplemental Material for further details.  
Note that since this model is triggered by the fermion density, the late-time bound objects can never exceed $n_\psi(z_\text{form})$ even locally. Since (by construction) this is comparable to the dwarf galaxy density, the greatest DM density attained on astrophysical scales, this feature does not disrupt the formation and stability of galactic structures.

\section{Outlook}
 We have demonstrated a mechanism by which fermionic DM well below the Tremaine--Gunn limit, down to the eV scale, can reproduce the dwarf galaxies that ordinarily set the bound. A forthcoming publication will address the resulting profiles and mass-radius relation for larger structures, where the interplay between gravity and the screened Yukawa interaction becomes crucial. The distinctive cosmological feature of this scenario is a sharp turn-on for the force at $z\sim 100$, which leaves the power spectrum untouched on observed scales while enhancing it below $\sim0.1\text{ Mpc}$. The baryonic consequences demand further investigation. For instance, the early nonlinear collapse may hasten star formation and the seeding of black holes, subject to constraints on the optical depth to reionization. This may be probed by 21-cm experiments, through the resulting shift in the onset of cosmic dawn. It may also be of possible relevance to the inference of overmassive black holes at high redshift, including in the little red dots~\cite{pacucci_jwst_2023,matthee_little_2024,greene_uncover_2024}. Notably, these signatures follow from the formation epoch alone and are independent of the trigger sector that realizes it.

%TC:ignore
{\bf Acknowledgement}.   We would like to thank Salvatore Bottaro, Konstantin Springmann, and Stefan Stelzl for valuable discussions. A.M. is supported by the Mortimer B. Zuckerman Foundation, and acknowledges the generous hospitality of IFAE Barcelona and the CERN Theory Department amidst geopolitical disruption to the writing of this work. J.B. thanks the Alexander Zaks scholarship for its support.
T.V.\ 
acknowledges support, in part, from the Binational Science Foundation (grant No. 2024160), the Israel Science
Foundation (grant No. 2803/25), the NSF-BSF (grant No. 2024823) and the PAZY Foundation (grant No. 429).
D.B. is supported by ERC grant ERC-2024-SYG 101167211 funded by the European Union, and acknowledges financial support from the Spanish Ministry of Science and Innovation (MICINN) through the Spanish State Research Agency, under Severo Ochoa Centres of Excellence Programme 2025-2029 (CEX2024001442-S).
IFAE is partially funded by the CERCA program of the Generalitat de Catalunya.
This publication is part of the R\&D\&i project PID2023-146686NB-C31 funded by MICIU/AEI/10.13039/501100011033/ and by ERDF/EU.
Views and opinions expressed are those of the author(s) only and do not necessarily reflect those of the European Union or the European Research Council Executive Agency. Neither the European Union nor the granting authority can be held responsible for them.

\bibliographystyle{apsrev4-2}
\bibliography{references}

%apsrev4-2.bst 2019-01-14 (MD) hand-edited version of apsrev4-1.bst
%Control: key (0)
%Control: author (72) initials jnrlst
%Control: editor formatted (1) identically to author
%Control: production of article title (-1) disabled
%Control: page (0) single
%Control: year (1) truncated
%Control: production of eprint (0) enabled
\begin{thebibliography}{81}%
\makeatletter
\providecommand \@ifxundefined [1]{%
 \@ifx{#1\undefined}
}%
\providecommand \@ifnum [1]{%
 \ifnum #1\expandafter \@firstoftwo
 \else \expandafter \@secondoftwo
 \fi
}%
\providecommand \@ifx [1]{%
 \ifx #1\expandafter \@firstoftwo
 \else \expandafter \@secondoftwo
 \fi
}%
\providecommand \natexlab [1]{#1}%
\providecommand \enquote  [1]{``#1''}%
\providecommand \bibnamefont  [1]{#1}%
\providecommand \bibfnamefont [1]{#1}%
\providecommand \citenamefont [1]{#1}%
\providecommand \href@noop [0]{\@secondoftwo}%
\providecommand \href [0]{\begingroup \@sanitize@url \@href}%
\providecommand \@href[1]{\@@startlink{#1}\@@href}%
\providecommand \@@href[1]{\endgroup#1\@@endlink}%
\providecommand \@sanitize@url [0]{\catcode `\\12\catcode `\$12\catcode `\&12\catcode `\#12\catcode `\^12\catcode `\_12\catcode `\%12\relax}%
\providecommand \@@startlink[1]{}%
\providecommand \@@endlink[0]{}%
\providecommand \url  [0]{\begingroup\@sanitize@url \@url }%
\providecommand \@url [1]{\endgroup\@href {#1}{\urlprefix }}%
\providecommand \urlprefix  [0]{URL }%
\providecommand \Eprint [0]{\href }%
\providecommand \doibase [0]{https://doi.org/}%
\providecommand \selectlanguage [0]{\@gobble}%
\providecommand \bibinfo  [0]{\@secondoftwo}%
\providecommand \bibfield  [0]{\@secondoftwo}%
\providecommand \translation [1]{[#1]}%
\providecommand \BibitemOpen [0]{}%
\providecommand \bibitemStop [0]{}%
\providecommand \bibitemNoStop [0]{.\EOS\space}%
\providecommand \EOS [0]{\spacefactor3000\relax}%
\providecommand \BibitemShut  [1]{\csname bibitem#1\endcsname}%
\let\auto@bib@innerbib\@empty
%</preamble>
\bibitem [{\citenamefont {Walker}(2013)}]{walker_dark_2013}%
  \BibitemOpen
  \bibfield  {author} {\bibinfo {author} {\bibfnamefont {M.}~\bibnamefont {Walker}},\ }in\ \href {https://doi.org/10.1007/978-94-007-5612-0_20} {\emph {\bibinfo {booktitle} {Planets, {{Stars}} and {{Stellar Systems}}: {{Volume}} 5: {{Galactic Structure}} and {{Stellar Populations}}}}},\ \bibinfo {editor} {edited by\ \bibinfo {editor} {\bibfnamefont {T.~D.}\ \bibnamefont {Oswalt}}\ and\ \bibinfo {editor} {\bibfnamefont {G.}~\bibnamefont {Gilmore}}}\ (\bibinfo  {publisher} {Springer Netherlands},\ \bibinfo {year} {2013})\ pp.\ \bibinfo {pages} {1039--1089}\BibitemShut {NoStop}%
\bibitem [{\citenamefont {Clowe}\ \emph {et~al.}(2006)\citenamefont {Clowe}, \citenamefont {Brada{\v c}}, \citenamefont {Gonzalez}, \citenamefont {Markevitch}, \citenamefont {Randall}, \citenamefont {Jones},\ and\ \citenamefont {Zaritsky}}]{clowe_direct_2006}%
  \BibitemOpen
  \bibfield  {author} {\bibinfo {author} {\bibfnamefont {D.}~\bibnamefont {Clowe}}, \bibinfo {author} {\bibfnamefont {M.}~\bibnamefont {Brada{\v c}}}, \bibinfo {author} {\bibfnamefont {A.~H.}\ \bibnamefont {Gonzalez}}, \bibinfo {author} {\bibfnamefont {M.}~\bibnamefont {Markevitch}}, \bibinfo {author} {\bibfnamefont {S.~W.}\ \bibnamefont {Randall}}, \bibinfo {author} {\bibfnamefont {C.}~\bibnamefont {Jones}},\ and\ \bibinfo {author} {\bibfnamefont {D.}~\bibnamefont {Zaritsky}},\ }\href {https://doi.org/10.1086/508162} {\bibfield  {journal} {\bibinfo  {journal} {ApJ}\ }\textbf {\bibinfo {volume} {648}},\ \bibinfo {pages} {L109} (\bibinfo {year} {2006})}\BibitemShut {NoStop}%
\bibitem [{\citenamefont {{de Martino}}\ \emph {et~al.}(2020)\citenamefont {{de Martino}}, \citenamefont {Chakrabarty}, \citenamefont {Cesare}, \citenamefont {Gallo}, \citenamefont {Ostorero},\ and\ \citenamefont {Diaferio}}]{de_martino_dark_2020}%
  \BibitemOpen
  \bibfield  {author} {\bibinfo {author} {\bibfnamefont {I.}~\bibnamefont {{de Martino}}}, \bibinfo {author} {\bibfnamefont {S.~S.}\ \bibnamefont {Chakrabarty}}, \bibinfo {author} {\bibfnamefont {V.}~\bibnamefont {Cesare}}, \bibinfo {author} {\bibfnamefont {A.}~\bibnamefont {Gallo}}, \bibinfo {author} {\bibfnamefont {L.}~\bibnamefont {Ostorero}},\ and\ \bibinfo {author} {\bibfnamefont {A.}~\bibnamefont {Diaferio}},\ }\href {https://doi.org/10.3390/universe6080107} {\bibfield  {journal} {\bibinfo  {journal} {Universe}\ }\textbf {\bibinfo {volume} {6}},\ \bibinfo {pages} {107} (\bibinfo {year} {2020})}\BibitemShut {NoStop}%
\bibitem [{\citenamefont {Tremaine}\ and\ \citenamefont {Gunn}(1979)}]{tremaine_dynamical_1979}%
  \BibitemOpen
  \bibfield  {author} {\bibinfo {author} {\bibfnamefont {S.}~\bibnamefont {Tremaine}}\ and\ \bibinfo {author} {\bibfnamefont {J.~E.}\ \bibnamefont {Gunn}},\ }\href {https://doi.org/10.1103/PhysRevLett.42.407} {\bibfield  {journal} {\bibinfo  {journal} {Phys. Rev. Lett.}\ }\textbf {\bibinfo {volume} {42}},\ \bibinfo {pages} {407} (\bibinfo {year} {1979})}\BibitemShut {NoStop}%
\bibitem [{\citenamefont {Madsen}(1991)}]{madsen_generalized_1991}%
  \BibitemOpen
  \bibfield  {author} {\bibinfo {author} {\bibfnamefont {J.}~\bibnamefont {Madsen}},\ }\href {https://doi.org/10.1103/PhysRevD.44.999} {\bibfield  {journal} {\bibinfo  {journal} {Phys. Rev. D}\ }\textbf {\bibinfo {volume} {44}},\ \bibinfo {pages} {999} (\bibinfo {year} {1991})}\BibitemShut {NoStop}%
\bibitem [{\citenamefont {Domcke}\ and\ \citenamefont {Urbano}(2015)}]{domcke_dwarf_2015}%
  \BibitemOpen
  \bibfield  {author} {\bibinfo {author} {\bibfnamefont {V.}~\bibnamefont {Domcke}}\ and\ \bibinfo {author} {\bibfnamefont {A.}~\bibnamefont {Urbano}},\ }\href {https://doi.org/10.1088/1475-7516/2015/01/002} {\bibfield  {journal} {\bibinfo  {journal} {J. Cosmol. Astropart. Phys.}\ }\textbf {\bibinfo {volume} {2015}}\bibfield  {number} {\bibinfo  {number} { (01)},\ \bibinfo {pages} {002}},\ }\Eprint {https://arxiv.org/abs/1409.3167} {arXiv:1409.3167 [astro-ph]} \BibitemShut {NoStop}%
\bibitem [{\citenamefont {Randall}\ \emph {et~al.}(2017)\citenamefont {Randall}, \citenamefont {Scholtz},\ and\ \citenamefont {Unwin}}]{randall_cores_2017}%
  \BibitemOpen
  \bibfield  {author} {\bibinfo {author} {\bibfnamefont {L.}~\bibnamefont {Randall}}, \bibinfo {author} {\bibfnamefont {J.}~\bibnamefont {Scholtz}},\ and\ \bibinfo {author} {\bibfnamefont {J.}~\bibnamefont {Unwin}},\ }\href {https://doi.org/10.1093/mnras/stx161} {\bibfield  {journal} {\bibinfo  {journal} {Monthly Notices of the Royal Astronomical Society}\ }\textbf {\bibinfo {volume} {467}},\ \bibinfo {pages} {1515} (\bibinfo {year} {2017})}\BibitemShut {NoStop}%
\bibitem [{\citenamefont {Bullock}\ and\ \citenamefont {{Boylan-Kolchin}}(2017)}]{bullock_small-scale_2017}%
  \BibitemOpen
  \bibfield  {author} {\bibinfo {author} {\bibfnamefont {J.~S.}\ \bibnamefont {Bullock}}\ and\ \bibinfo {author} {\bibfnamefont {M.}~\bibnamefont {{Boylan-Kolchin}}},\ }\href {https://doi.org/10.1146/annurev-astro-091916-055313} {\bibfield  {journal} {\bibinfo  {journal} {Annual Review of Astronomy and Astrophysics}\ }\textbf {\bibinfo {volume} {55}},\ \bibinfo {pages} {343} (\bibinfo {year} {2017})}\BibitemShut {NoStop}%
\bibitem [{\citenamefont {Tulin}\ and\ \citenamefont {Yu}(2018)}]{tulin_dark_2018}%
  \BibitemOpen
  \bibfield  {author} {\bibinfo {author} {\bibfnamefont {S.}~\bibnamefont {Tulin}}\ and\ \bibinfo {author} {\bibfnamefont {H.-B.}\ \bibnamefont {Yu}},\ }\href {https://doi.org/10.1016/j.physrep.2017.11.004} {\bibfield  {journal} {\bibinfo  {journal} {Physics Reports}\ }\bibinfo {series} {Dark Matter Self-Interactions and Small Scale Structure},\ \textbf {\bibinfo {volume} {730}},\ \bibinfo {pages} {1} (\bibinfo {year} {2018})}\BibitemShut {NoStop}%
\bibitem [{\citenamefont {Spergel}\ and\ \citenamefont {Steinhardt}(2000)}]{spergel_observational_2000}%
  \BibitemOpen
  \bibfield  {author} {\bibinfo {author} {\bibfnamefont {D.~N.}\ \bibnamefont {Spergel}}\ and\ \bibinfo {author} {\bibfnamefont {P.~J.}\ \bibnamefont {Steinhardt}},\ }\href {https://doi.org/10.1103/PhysRevLett.84.3760} {\bibfield  {journal} {\bibinfo  {journal} {Phys. Rev. Lett.}\ }\textbf {\bibinfo {volume} {84}},\ \bibinfo {pages} {3760} (\bibinfo {year} {2000})}\BibitemShut {NoStop}%
\bibitem [{\citenamefont {Loeb}(2011)}]{loeb_cores_2011}%
  \BibitemOpen
  \bibfield  {author} {\bibinfo {author} {\bibfnamefont {A.}~\bibnamefont {Loeb}},\ }\bibfield  {journal} {\bibinfo  {journal} {Phys. Rev. Lett.}\ }\textbf {\bibinfo {volume} {106}},\ \href {https://doi.org/10.1103/PhysRevLett.106.171302} {10.1103/PhysRevLett.106.171302} (\bibinfo {year} {2011})\BibitemShut {NoStop}%
\bibitem [{\citenamefont {Kaplinghat}\ \emph {et~al.}(2016)\citenamefont {Kaplinghat}, \citenamefont {Tulin},\ and\ \citenamefont {Yu}}]{kaplinghat_dark_2016}%
  \BibitemOpen
  \bibfield  {author} {\bibinfo {author} {\bibfnamefont {M.}~\bibnamefont {Kaplinghat}}, \bibinfo {author} {\bibfnamefont {S.}~\bibnamefont {Tulin}},\ and\ \bibinfo {author} {\bibfnamefont {H.-B.}\ \bibnamefont {Yu}},\ }\href {https://doi.org/10.1103/PhysRevLett.116.041302} {\bibfield  {journal} {\bibinfo  {journal} {Phys. Rev. Lett.}\ }\textbf {\bibinfo {volume} {116}},\ \bibinfo {pages} {041302} (\bibinfo {year} {2016})}\BibitemShut {NoStop}%
\bibitem [{\citenamefont {Hu}\ \emph {et~al.}(2000)\citenamefont {Hu}, \citenamefont {Barkana},\ and\ \citenamefont {Gruzinov}}]{hu_fuzzy_2000}%
  \BibitemOpen
  \bibfield  {author} {\bibinfo {author} {\bibfnamefont {W.}~\bibnamefont {Hu}}, \bibinfo {author} {\bibfnamefont {R.}~\bibnamefont {Barkana}},\ and\ \bibinfo {author} {\bibfnamefont {A.}~\bibnamefont {Gruzinov}},\ }\href {https://doi.org/10.1103/PhysRevLett.85.1158} {\bibfield  {journal} {\bibinfo  {journal} {Phys. Rev. Lett.}\ }\textbf {\bibinfo {volume} {85}},\ \bibinfo {pages} {1158} (\bibinfo {year} {2000})}\BibitemShut {NoStop}%
\bibitem [{\citenamefont {Schive}\ \emph {et~al.}(2014)\citenamefont {Schive}, \citenamefont {Chiueh},\ and\ \citenamefont {Broadhurst}}]{schive_cosmic_2014}%
  \BibitemOpen
  \bibfield  {author} {\bibinfo {author} {\bibfnamefont {H.-Y.}\ \bibnamefont {Schive}}, \bibinfo {author} {\bibfnamefont {T.}~\bibnamefont {Chiueh}},\ and\ \bibinfo {author} {\bibfnamefont {T.}~\bibnamefont {Broadhurst}},\ }\href {https://doi.org/10.1038/nphys2996} {\bibfield  {journal} {\bibinfo  {journal} {Nature Phys}\ }\textbf {\bibinfo {volume} {10}},\ \bibinfo {pages} {496} (\bibinfo {year} {2014})}\BibitemShut {NoStop}%
\bibitem [{\citenamefont {Hui}\ \emph {et~al.}(2017)\citenamefont {Hui}, \citenamefont {Ostriker}, \citenamefont {Tremaine},\ and\ \citenamefont {Witten}}]{hui_ultralight_2017}%
  \BibitemOpen
  \bibfield  {author} {\bibinfo {author} {\bibfnamefont {L.}~\bibnamefont {Hui}}, \bibinfo {author} {\bibfnamefont {J.~P.}\ \bibnamefont {Ostriker}}, \bibinfo {author} {\bibfnamefont {S.}~\bibnamefont {Tremaine}},\ and\ \bibinfo {author} {\bibfnamefont {E.}~\bibnamefont {Witten}},\ }\href {https://doi.org/10.1103/PhysRevD.95.043541} {\bibfield  {journal} {\bibinfo  {journal} {Phys. Rev. D}\ }\textbf {\bibinfo {volume} {95}},\ \bibinfo {pages} {043541} (\bibinfo {year} {2017})},\ \Eprint {https://arxiv.org/abs/1610.08297} {arXiv:1610.08297 [astro-ph]} \BibitemShut {NoStop}%
\bibitem [{\citenamefont {Ferreira}(2021)}]{ferreira_ultra-light_2021}%
  \BibitemOpen
  \bibfield  {author} {\bibinfo {author} {\bibfnamefont {E.~G.~M.}\ \bibnamefont {Ferreira}},\ }\href {https://doi.org/10.1007/s00159-021-00135-6} {\bibfield  {journal} {\bibinfo  {journal} {Astron Astrophys Rev}\ }\textbf {\bibinfo {volume} {29}},\ \bibinfo {pages} {7} (\bibinfo {year} {2021})}\BibitemShut {NoStop}%
\bibitem [{\citenamefont {Antypas}\ \emph {et~al.}(2022)\citenamefont {Antypas}, \citenamefont {Banerjee}, \citenamefont {Bartram}, \citenamefont {Baryakhtar}, \citenamefont {Betz}, \citenamefont {Bollinger}, \citenamefont {Boutan}, \citenamefont {Bowring}, \citenamefont {Budker}, \citenamefont {Carney}, \citenamefont {Carosi}, \citenamefont {Chaudhuri}, \citenamefont {Cheong}, \citenamefont {Chou}, \citenamefont {Chowdhury}, \citenamefont {Co}, \citenamefont {{L{\'o}pez-Urrutia}}, \citenamefont {Demarteau}, \citenamefont {DePorzio}, \citenamefont {Derbin}, \citenamefont {Deshpande}, \citenamefont {Chowdhury}, \citenamefont {Luzio}, \citenamefont {{Diaz-Morcillo}}, \citenamefont {Doyle}, \citenamefont {{Drlica-Wagner}}, \citenamefont {Droster}, \citenamefont {Du}, \citenamefont {D{\"o}brich}, \citenamefont {Eby}, \citenamefont {Essig}, \citenamefont {Farren}, \citenamefont {Figueroa}, \citenamefont {Fry}, \citenamefont {Gardner}, \citenamefont {Geraci}, \citenamefont {Ghalsasi}, \citenamefont {Ghosh},
  \citenamefont {Giannotti}, \citenamefont {Gimeno}, \citenamefont {Griffin}, \citenamefont {Grin}, \citenamefont {Grin}, \citenamefont {Grote}, \citenamefont {Gundlach}, \citenamefont {Guzzetti}, \citenamefont {Hanneke}, \citenamefont {Harnik}, \citenamefont {Henning}, \citenamefont {Irsic}, \citenamefont {Jackson}, \citenamefont {Kimball}, \citenamefont {Jaeckel}, \citenamefont {Kagan}, \citenamefont {Kedar}, \citenamefont {Khatiwada}, \citenamefont {Knirck}, \citenamefont {Kolkowitz}, \citenamefont {Kovachy}, \citenamefont {Kuenstner}, \citenamefont {Lasner}, \citenamefont {Leder}, \citenamefont {Lehnert}, \citenamefont {Leibrandt}, \citenamefont {Lentz}, \citenamefont {Lewis}, \citenamefont {Liu}, \citenamefont {Manley}, \citenamefont {Maruyama}, \citenamefont {Millar}, \citenamefont {Muratova}, \citenamefont {Musoke}, \citenamefont {Nagaitsev}, \citenamefont {Noroozian}, \citenamefont {O'Hare}, \citenamefont {Ouellet}, \citenamefont {Pappas}, \citenamefont {Peik}, \citenamefont {Perez}, \citenamefont
  {Phipps}, \citenamefont {Rapidis}, \citenamefont {Robinson}, \citenamefont {Robles}, \citenamefont {Rogers}, \citenamefont {Rudolph}, \citenamefont {Rybka}, \citenamefont {Safdari}, \citenamefont {Safdari}, \citenamefont {Safronova}, \citenamefont {Salemi}, \citenamefont {Schmidt}, \citenamefont {Schumm}, \citenamefont {Schwartzman}, \citenamefont {Shu}, \citenamefont {Simanovskaia}, \citenamefont {Singh}, \citenamefont {Singh}, \citenamefont {Smith}, \citenamefont {Snow}, \citenamefont {Stadnik}, \citenamefont {Sun}, \citenamefont {Sushkov}, \citenamefont {Tait}, \citenamefont {Takhistov}, \citenamefont {Tanner}, \citenamefont {Temples}, \citenamefont {Thirolf}, \citenamefont {Thomas}, \citenamefont {Tobar}, \citenamefont {Tretiak}, \citenamefont {Tsai}, \citenamefont {Tyson}, \citenamefont {Vandegar}, \citenamefont {Vermeulen}, \citenamefont {Visinelli}, \citenamefont {Vitagliano}, \citenamefont {Wang}, \citenamefont {Wilson}, \citenamefont {Winslow}, \citenamefont {Withington}, \citenamefont {Wooten},
  \citenamefont {Yang}, \citenamefont {Ye}, \citenamefont {Young}, \citenamefont {Yu}, \citenamefont {Zaheer}, \citenamefont {Zelevinsky}, \citenamefont {Zhao},\ and\ \citenamefont {Zhou}}]{antypas_new_2022}%
  \BibitemOpen
  \bibfield  {author} {\bibinfo {author} {\bibfnamefont {D.}~\bibnamefont {Antypas}}, \bibinfo {author} {\bibfnamefont {A.}~\bibnamefont {Banerjee}}, \bibinfo {author} {\bibfnamefont {C.}~\bibnamefont {Bartram}}, \bibinfo {author} {\bibfnamefont {M.}~\bibnamefont {Baryakhtar}}, \bibinfo {author} {\bibfnamefont {J.}~\bibnamefont {Betz}}, \bibinfo {author} {\bibfnamefont {J.~J.}\ \bibnamefont {Bollinger}}, \bibinfo {author} {\bibfnamefont {C.}~\bibnamefont {Boutan}}, \bibinfo {author} {\bibfnamefont {D.}~\bibnamefont {Bowring}}, \bibinfo {author} {\bibfnamefont {D.}~\bibnamefont {Budker}}, \bibinfo {author} {\bibfnamefont {D.}~\bibnamefont {Carney}}, \bibinfo {author} {\bibfnamefont {G.}~\bibnamefont {Carosi}}, \bibinfo {author} {\bibfnamefont {S.}~\bibnamefont {Chaudhuri}}, \bibinfo {author} {\bibfnamefont {S.}~\bibnamefont {Cheong}}, \bibinfo {author} {\bibfnamefont {A.}~\bibnamefont {Chou}}, \bibinfo {author} {\bibfnamefont {M.~D.}\ \bibnamefont {Chowdhury}}, \bibinfo {author} {\bibfnamefont {R.~T.}\
  \bibnamefont {Co}}, \bibinfo {author} {\bibfnamefont {J.~R.~C.}\ \bibnamefont {{L{\'o}pez-Urrutia}}}, \bibinfo {author} {\bibfnamefont {M.}~\bibnamefont {Demarteau}}, \bibinfo {author} {\bibfnamefont {N.}~\bibnamefont {DePorzio}}, \bibinfo {author} {\bibfnamefont {A.~V.}\ \bibnamefont {Derbin}}, \bibinfo {author} {\bibfnamefont {T.}~\bibnamefont {Deshpande}}, \bibinfo {author} {\bibfnamefont {M.~D.}\ \bibnamefont {Chowdhury}}, \bibinfo {author} {\bibfnamefont {L.~D.}\ \bibnamefont {Luzio}}, \bibinfo {author} {\bibfnamefont {A.}~\bibnamefont {{Diaz-Morcillo}}}, \bibinfo {author} {\bibfnamefont {J.~M.}\ \bibnamefont {Doyle}}, \bibinfo {author} {\bibfnamefont {A.}~\bibnamefont {{Drlica-Wagner}}}, \bibinfo {author} {\bibfnamefont {A.}~\bibnamefont {Droster}}, \bibinfo {author} {\bibfnamefont {N.}~\bibnamefont {Du}}, \bibinfo {author} {\bibfnamefont {B.}~\bibnamefont {D{\"o}brich}}, \bibinfo {author} {\bibfnamefont {J.}~\bibnamefont {Eby}}, \bibinfo {author} {\bibfnamefont {R.}~\bibnamefont {Essig}}, \bibinfo
  {author} {\bibfnamefont {G.~S.}\ \bibnamefont {Farren}}, \bibinfo {author} {\bibfnamefont {N.~L.}\ \bibnamefont {Figueroa}}, \bibinfo {author} {\bibfnamefont {J.~T.}\ \bibnamefont {Fry}}, \bibinfo {author} {\bibfnamefont {S.}~\bibnamefont {Gardner}}, \bibinfo {author} {\bibfnamefont {A.~A.}\ \bibnamefont {Geraci}}, \bibinfo {author} {\bibfnamefont {A.}~\bibnamefont {Ghalsasi}}, \bibinfo {author} {\bibfnamefont {S.}~\bibnamefont {Ghosh}}, \bibinfo {author} {\bibfnamefont {M.}~\bibnamefont {Giannotti}}, \bibinfo {author} {\bibfnamefont {B.}~\bibnamefont {Gimeno}}, \bibinfo {author} {\bibfnamefont {S.~M.}\ \bibnamefont {Griffin}}, \bibinfo {author} {\bibfnamefont {D.}~\bibnamefont {Grin}}, \bibinfo {author} {\bibfnamefont {D.}~\bibnamefont {Grin}}, \bibinfo {author} {\bibfnamefont {H.}~\bibnamefont {Grote}}, \bibinfo {author} {\bibfnamefont {J.~H.}\ \bibnamefont {Gundlach}}, \bibinfo {author} {\bibfnamefont {M.}~\bibnamefont {Guzzetti}}, \bibinfo {author} {\bibfnamefont {D.}~\bibnamefont {Hanneke}}, \bibinfo
  {author} {\bibfnamefont {R.}~\bibnamefont {Harnik}}, \bibinfo {author} {\bibfnamefont {R.}~\bibnamefont {Henning}}, \bibinfo {author} {\bibfnamefont {V.}~\bibnamefont {Irsic}}, \bibinfo {author} {\bibfnamefont {H.}~\bibnamefont {Jackson}}, \bibinfo {author} {\bibfnamefont {D.~F.~J.}\ \bibnamefont {Kimball}}, \bibinfo {author} {\bibfnamefont {J.}~\bibnamefont {Jaeckel}}, \bibinfo {author} {\bibfnamefont {M.}~\bibnamefont {Kagan}}, \bibinfo {author} {\bibfnamefont {D.}~\bibnamefont {Kedar}}, \bibinfo {author} {\bibfnamefont {R.}~\bibnamefont {Khatiwada}}, \bibinfo {author} {\bibfnamefont {S.}~\bibnamefont {Knirck}}, \bibinfo {author} {\bibfnamefont {S.}~\bibnamefont {Kolkowitz}}, \bibinfo {author} {\bibfnamefont {T.}~\bibnamefont {Kovachy}}, \bibinfo {author} {\bibfnamefont {S.~E.}\ \bibnamefont {Kuenstner}}, \bibinfo {author} {\bibfnamefont {Z.}~\bibnamefont {Lasner}}, \bibinfo {author} {\bibfnamefont {A.~F.}\ \bibnamefont {Leder}}, \bibinfo {author} {\bibfnamefont {R.}~\bibnamefont {Lehnert}}, \bibinfo
  {author} {\bibfnamefont {D.~R.}\ \bibnamefont {Leibrandt}}, \bibinfo {author} {\bibfnamefont {E.}~\bibnamefont {Lentz}}, \bibinfo {author} {\bibfnamefont {S.~M.}\ \bibnamefont {Lewis}}, \bibinfo {author} {\bibfnamefont {Z.}~\bibnamefont {Liu}}, \bibinfo {author} {\bibfnamefont {J.}~\bibnamefont {Manley}}, \bibinfo {author} {\bibfnamefont {R.~H.}\ \bibnamefont {Maruyama}}, \bibinfo {author} {\bibfnamefont {A.~J.}\ \bibnamefont {Millar}}, \bibinfo {author} {\bibfnamefont {V.~N.}\ \bibnamefont {Muratova}}, \bibinfo {author} {\bibfnamefont {N.}~\bibnamefont {Musoke}}, \bibinfo {author} {\bibfnamefont {S.}~\bibnamefont {Nagaitsev}}, \bibinfo {author} {\bibfnamefont {O.}~\bibnamefont {Noroozian}}, \bibinfo {author} {\bibfnamefont {C.~A.~J.}\ \bibnamefont {O'Hare}}, \bibinfo {author} {\bibfnamefont {J.~L.}\ \bibnamefont {Ouellet}}, \bibinfo {author} {\bibfnamefont {K.~M.~W.}\ \bibnamefont {Pappas}}, \bibinfo {author} {\bibfnamefont {E.}~\bibnamefont {Peik}}, \bibinfo {author} {\bibfnamefont {G.}~\bibnamefont
  {Perez}}, \bibinfo {author} {\bibfnamefont {A.}~\bibnamefont {Phipps}}, \bibinfo {author} {\bibfnamefont {N.~M.}\ \bibnamefont {Rapidis}}, \bibinfo {author} {\bibfnamefont {J.~M.}\ \bibnamefont {Robinson}}, \bibinfo {author} {\bibfnamefont {V.~H.}\ \bibnamefont {Robles}}, \bibinfo {author} {\bibfnamefont {K.~K.}\ \bibnamefont {Rogers}}, \bibinfo {author} {\bibfnamefont {J.}~\bibnamefont {Rudolph}}, \bibinfo {author} {\bibfnamefont {G.}~\bibnamefont {Rybka}}, \bibinfo {author} {\bibfnamefont {M.}~\bibnamefont {Safdari}}, \bibinfo {author} {\bibfnamefont {M.}~\bibnamefont {Safdari}}, \bibinfo {author} {\bibfnamefont {M.~S.}\ \bibnamefont {Safronova}}, \bibinfo {author} {\bibfnamefont {C.~P.}\ \bibnamefont {Salemi}}, \bibinfo {author} {\bibfnamefont {P.~O.}\ \bibnamefont {Schmidt}}, \bibinfo {author} {\bibfnamefont {T.}~\bibnamefont {Schumm}}, \bibinfo {author} {\bibfnamefont {A.}~\bibnamefont {Schwartzman}}, \bibinfo {author} {\bibfnamefont {J.}~\bibnamefont {Shu}}, \bibinfo {author} {\bibfnamefont
  {M.}~\bibnamefont {Simanovskaia}}, \bibinfo {author} {\bibfnamefont {J.}~\bibnamefont {Singh}}, \bibinfo {author} {\bibfnamefont {S.}~\bibnamefont {Singh}}, \bibinfo {author} {\bibfnamefont {M.~S.}\ \bibnamefont {Smith}}, \bibinfo {author} {\bibfnamefont {W.~M.}\ \bibnamefont {Snow}}, \bibinfo {author} {\bibfnamefont {Y.~V.}\ \bibnamefont {Stadnik}}, \bibinfo {author} {\bibfnamefont {C.}~\bibnamefont {Sun}}, \bibinfo {author} {\bibfnamefont {A.~O.}\ \bibnamefont {Sushkov}}, \bibinfo {author} {\bibfnamefont {T.~M.~P.}\ \bibnamefont {Tait}}, \bibinfo {author} {\bibfnamefont {V.}~\bibnamefont {Takhistov}}, \bibinfo {author} {\bibfnamefont {D.~B.}\ \bibnamefont {Tanner}}, \bibinfo {author} {\bibfnamefont {D.~J.}\ \bibnamefont {Temples}}, \bibinfo {author} {\bibfnamefont {P.~G.}\ \bibnamefont {Thirolf}}, \bibinfo {author} {\bibfnamefont {J.~H.}\ \bibnamefont {Thomas}}, \bibinfo {author} {\bibfnamefont {M.~E.}\ \bibnamefont {Tobar}}, \bibinfo {author} {\bibfnamefont {O.}~\bibnamefont {Tretiak}}, \bibinfo {author}
  {\bibfnamefont {Y.-D.}\ \bibnamefont {Tsai}}, \bibinfo {author} {\bibfnamefont {J.~A.}\ \bibnamefont {Tyson}}, \bibinfo {author} {\bibfnamefont {M.}~\bibnamefont {Vandegar}}, \bibinfo {author} {\bibfnamefont {S.}~\bibnamefont {Vermeulen}}, \bibinfo {author} {\bibfnamefont {L.}~\bibnamefont {Visinelli}}, \bibinfo {author} {\bibfnamefont {E.}~\bibnamefont {Vitagliano}}, \bibinfo {author} {\bibfnamefont {Z.}~\bibnamefont {Wang}}, \bibinfo {author} {\bibfnamefont {D.~J.}\ \bibnamefont {Wilson}}, \bibinfo {author} {\bibfnamefont {L.}~\bibnamefont {Winslow}}, \bibinfo {author} {\bibfnamefont {S.}~\bibnamefont {Withington}}, \bibinfo {author} {\bibfnamefont {M.}~\bibnamefont {Wooten}}, \bibinfo {author} {\bibfnamefont {J.}~\bibnamefont {Yang}}, \bibinfo {author} {\bibfnamefont {J.}~\bibnamefont {Ye}}, \bibinfo {author} {\bibfnamefont {B.~A.}\ \bibnamefont {Young}}, \bibinfo {author} {\bibfnamefont {F.}~\bibnamefont {Yu}}, \bibinfo {author} {\bibfnamefont {M.~H.}\ \bibnamefont {Zaheer}}, \bibinfo {author}
  {\bibfnamefont {T.}~\bibnamefont {Zelevinsky}}, \bibinfo {author} {\bibfnamefont {Y.}~\bibnamefont {Zhao}},\ and\ \bibinfo {author} {\bibfnamefont {K.}~\bibnamefont {Zhou}},\ }\href {https://doi.org/10.48550/arXiv.2203.14915} {\bibinfo {title} {New {{Horizons}}: {{Scalar}} and {{Vector Ultralight Dark Matter}}}} (\bibinfo {year} {2022}),\ \Eprint {https://arxiv.org/abs/2203.14915} {arXiv:2203.14915 [hep-ex]} \BibitemShut {NoStop}%
\bibitem [{\citenamefont {Alexander}\ and\ \citenamefont {Cormack}(2017)}]{alexander_gravitationally_2017}%
  \BibitemOpen
  \bibfield  {author} {\bibinfo {author} {\bibfnamefont {S.}~\bibnamefont {Alexander}}\ and\ \bibinfo {author} {\bibfnamefont {S.}~\bibnamefont {Cormack}},\ }\href {https://doi.org/10.1088/1475-7516/2017/04/005} {\bibfield  {journal} {\bibinfo  {journal} {J. Cosmol. Astropart. Phys.}\ }\textbf {\bibinfo {volume} {2017}}\bibfield  {number} {\bibinfo  {number} { (04)},\ \bibinfo {pages} {005}},\ }\Eprint {https://arxiv.org/abs/1607.08621} {arXiv:1607.08621 [astro-ph]} \BibitemShut {NoStop}%
\bibitem [{\citenamefont {Alexander}\ \emph {et~al.}(2021)\citenamefont {Alexander}, \citenamefont {McDonough},\ and\ \citenamefont {Spergel}}]{alexander_strongly-interacting_2021}%
  \BibitemOpen
  \bibfield  {author} {\bibinfo {author} {\bibfnamefont {S.}~\bibnamefont {Alexander}}, \bibinfo {author} {\bibfnamefont {E.}~\bibnamefont {McDonough}},\ and\ \bibinfo {author} {\bibfnamefont {D.~N.}\ \bibnamefont {Spergel}},\ }\href {https://doi.org/10.1016/j.physletb.2021.136653} {\bibfield  {journal} {\bibinfo  {journal} {Physics Letters B}\ }\textbf {\bibinfo {volume} {822}},\ \bibinfo {pages} {136653} (\bibinfo {year} {2021})}\BibitemShut {NoStop}%
\bibitem [{\citenamefont {Garani}\ \emph {et~al.}(2022)\citenamefont {Garani}, \citenamefont {Tytgat},\ and\ \citenamefont {Vandecasteele}}]{garani_condensed_2022}%
  \BibitemOpen
  \bibfield  {author} {\bibinfo {author} {\bibfnamefont {R.}~\bibnamefont {Garani}}, \bibinfo {author} {\bibfnamefont {M.~H.~G.}\ \bibnamefont {Tytgat}},\ and\ \bibinfo {author} {\bibfnamefont {J.}~\bibnamefont {Vandecasteele}},\ }\href {https://doi.org/10.1103/PhysRevD.106.116003} {\bibfield  {journal} {\bibinfo  {journal} {Phys. Rev. D}\ }\textbf {\bibinfo {volume} {106}},\ \bibinfo {pages} {116003} (\bibinfo {year} {2022})},\ \Eprint {https://arxiv.org/abs/2207.06928} {arXiv:2207.06928 [astro-ph]} \BibitemShut {NoStop}%
\bibitem [{\citenamefont {Davoudiasl}\ \emph {et~al.}(2021)\citenamefont {Davoudiasl}, \citenamefont {Denton},\ and\ \citenamefont {McGady}}]{davoudiasl_ultralight_2021}%
  \BibitemOpen
  \bibfield  {author} {\bibinfo {author} {\bibfnamefont {H.}~\bibnamefont {Davoudiasl}}, \bibinfo {author} {\bibfnamefont {P.~B.}\ \bibnamefont {Denton}},\ and\ \bibinfo {author} {\bibfnamefont {D.~A.}\ \bibnamefont {McGady}},\ }\href {https://doi.org/10.1103/PhysRevD.103.055014} {\bibfield  {journal} {\bibinfo  {journal} {Phys. Rev. D}\ }\textbf {\bibinfo {volume} {103}},\ \bibinfo {pages} {055014} (\bibinfo {year} {2021})}\BibitemShut {NoStop}%
\bibitem [{\citenamefont {Boubekeur}\ and\ \citenamefont {Profumo}(2023)}]{boubekeur_tremaine-gunn_2023}%
  \BibitemOpen
  \bibfield  {author} {\bibinfo {author} {\bibfnamefont {L.}~\bibnamefont {Boubekeur}}\ and\ \bibinfo {author} {\bibfnamefont {S.}~\bibnamefont {Profumo}},\ }\href {https://doi.org/10.1103/PhysRevD.107.103535} {\bibfield  {journal} {\bibinfo  {journal} {Phys. Rev. D}\ }\textbf {\bibinfo {volume} {107}},\ \bibinfo {pages} {103535} (\bibinfo {year} {2023})}\BibitemShut {NoStop}%
\bibitem [{\citenamefont {Cline}\ \emph {et~al.}(2026)\citenamefont {Cline}, \citenamefont {Herrera},\ and\ \citenamefont {Roux}}]{cline_neutrinos_2026}%
  \BibitemOpen
  \bibfield  {author} {\bibinfo {author} {\bibfnamefont {J.~M.}\ \bibnamefont {Cline}}, \bibinfo {author} {\bibfnamefont {G.}~\bibnamefont {Herrera}},\ and\ \bibinfo {author} {\bibfnamefont {J.-S.}\ \bibnamefont {Roux}},\ }\href {https://doi.org/10.48550/arXiv.2603.28859} {\bibinfo {title} {Neutrinos as {{Dark Matter}}}} (\bibinfo {year} {2026}),\ \Eprint {https://arxiv.org/abs/2603.28859} {arXiv:2603.28859 [hep-ph]} \BibitemShut {NoStop}%
\bibitem [{\citenamefont {Frieman}\ and\ \citenamefont {Gradwohl}(1991)}]{frieman_dark_1991}%
  \BibitemOpen
  \bibfield  {author} {\bibinfo {author} {\bibfnamefont {J.~A.}\ \bibnamefont {Frieman}}\ and\ \bibinfo {author} {\bibfnamefont {B.-A.}\ \bibnamefont {Gradwohl}},\ }\href {https://doi.org/10.1103/PhysRevLett.67.2926} {\bibfield  {journal} {\bibinfo  {journal} {Phys. Rev. Lett.}\ }\textbf {\bibinfo {volume} {67}},\ \bibinfo {pages} {2926} (\bibinfo {year} {1991})}\BibitemShut {NoStop}%
\bibitem [{\citenamefont {Gubser}\ and\ \citenamefont {Peebles}(2004)}]{gubser_cosmology_2004}%
  \BibitemOpen
  \bibfield  {author} {\bibinfo {author} {\bibfnamefont {S.~S.}\ \bibnamefont {Gubser}}\ and\ \bibinfo {author} {\bibfnamefont {P.~J.~E.}\ \bibnamefont {Peebles}},\ }\href {https://doi.org/10.1103/PhysRevD.70.123511} {\bibfield  {journal} {\bibinfo  {journal} {Phys. Rev. D}\ }\textbf {\bibinfo {volume} {70}},\ \bibinfo {pages} {123511} (\bibinfo {year} {2004})}\BibitemShut {NoStop}%
\bibitem [{\citenamefont {Nusser}\ \emph {et~al.}(2005)\citenamefont {Nusser}, \citenamefont {Gubser},\ and\ \citenamefont {Peebles}}]{nusser_structure_2005}%
  \BibitemOpen
  \bibfield  {author} {\bibinfo {author} {\bibfnamefont {A.}~\bibnamefont {Nusser}}, \bibinfo {author} {\bibfnamefont {S.~S.}\ \bibnamefont {Gubser}},\ and\ \bibinfo {author} {\bibfnamefont {P.~J.~E.}\ \bibnamefont {Peebles}},\ }\href {https://doi.org/10.1103/PhysRevD.71.083505} {\bibfield  {journal} {\bibinfo  {journal} {Phys. Rev. D}\ }\textbf {\bibinfo {volume} {71}},\ \bibinfo {pages} {083505} (\bibinfo {year} {2005})}\BibitemShut {NoStop}%
\bibitem [{\citenamefont {Graham}\ \emph {et~al.}(2025)\citenamefont {Graham}, \citenamefont {Ramani}, \citenamefont {Simon},\ and\ \citenamefont {Tanin}}]{graham_cosmological_2025}%
  \BibitemOpen
  \bibfield  {author} {\bibinfo {author} {\bibfnamefont {P.~W.}\ \bibnamefont {Graham}}, \bibinfo {author} {\bibfnamefont {H.}~\bibnamefont {Ramani}}, \bibinfo {author} {\bibfnamefont {O.}~\bibnamefont {Simon}},\ and\ \bibinfo {author} {\bibfnamefont {E.~H.}\ \bibnamefont {Tanin}},\ }\href {https://doi.org/10.48550/arXiv.2511.09614} {\bibinfo {title} {Cosmological {{Limits}} on {{Strong Dark Forces}}}} (\bibinfo {year} {2025}),\ \Eprint {https://arxiv.org/abs/2511.09614} {arXiv:2511.09614 [hep-ph]} \BibitemShut {NoStop}%
\bibitem [{\citenamefont {Gor'Kov}\ and\ \citenamefont {{Melik-Barkhudarov}}(1961)}]{gorkov_contribution_1961}%
  \BibitemOpen
  \bibfield  {author} {\bibinfo {author} {\bibfnamefont {L.~P.}\ \bibnamefont {Gor'Kov}}\ and\ \bibinfo {author} {\bibfnamefont {T.~K.}\ \bibnamefont {{Melik-Barkhudarov}}},\ }\href@noop {} {\bibfield  {journal} {\bibinfo  {journal} {Soviet Physics JETP}\ }\textbf {\bibinfo {volume} {13}},\ \bibinfo {pages} {1018} (\bibinfo {year} {1961})}\BibitemShut {NoStop}%
\bibitem [{\citenamefont {Kapusta}(2004)}]{kapusta_neutrino_2004}%
  \BibitemOpen
  \bibfield  {author} {\bibinfo {author} {\bibfnamefont {J.~I.}\ \bibnamefont {Kapusta}},\ }\href {https://doi.org/10.1103/PhysRevLett.93.251801} {\bibfield  {journal} {\bibinfo  {journal} {Phys. Rev. Lett.}\ }\textbf {\bibinfo {volume} {93}},\ \bibinfo {pages} {251801} (\bibinfo {year} {2004})},\ \Eprint {https://arxiv.org/abs/hep-th/0407164} {arXiv:hep-th/0407164} \BibitemShut {NoStop}%
\bibitem [{\citenamefont {Carena}\ \emph {et~al.}(2022)\citenamefont {Carena}, \citenamefont {Coyle}, \citenamefont {Li}, \citenamefont {McDermott},\ and\ \citenamefont {Tsai}}]{carena_cosmologically_2022}%
  \BibitemOpen
  \bibfield  {author} {\bibinfo {author} {\bibfnamefont {M.}~\bibnamefont {Carena}}, \bibinfo {author} {\bibfnamefont {N.~M.}\ \bibnamefont {Coyle}}, \bibinfo {author} {\bibfnamefont {Y.-Y.}\ \bibnamefont {Li}}, \bibinfo {author} {\bibfnamefont {S.~D.}\ \bibnamefont {McDermott}},\ and\ \bibinfo {author} {\bibfnamefont {Y.}~\bibnamefont {Tsai}},\ }\href {https://doi.org/10.1103/PhysRevD.106.083016} {\bibfield  {journal} {\bibinfo  {journal} {Phys. Rev. D}\ }\textbf {\bibinfo {volume} {106}},\ \bibinfo {pages} {083016} (\bibinfo {year} {2022})}\BibitemShut {NoStop}%
\bibitem [{\citenamefont {Dalcanton}\ and\ \citenamefont {Hogan}(2001)}]{dalcanton_halo_2001}%
  \BibitemOpen
  \bibfield  {author} {\bibinfo {author} {\bibfnamefont {J.~J.}\ \bibnamefont {Dalcanton}}\ and\ \bibinfo {author} {\bibfnamefont {C.~J.}\ \bibnamefont {Hogan}},\ }\href {https://doi.org/10.1086/323207} {\bibfield  {journal} {\bibinfo  {journal} {ApJ}\ }\textbf {\bibinfo {volume} {561}},\ \bibinfo {pages} {35} (\bibinfo {year} {2001})}\BibitemShut {NoStop}%
\bibitem [{\citenamefont {Boyarsky}\ \emph {et~al.}(2009)\citenamefont {Boyarsky}, \citenamefont {Ruchayskiy},\ and\ \citenamefont {Iakubovskyi}}]{boyarsky_lower_2009}%
  \BibitemOpen
  \bibfield  {author} {\bibinfo {author} {\bibfnamefont {A.}~\bibnamefont {Boyarsky}}, \bibinfo {author} {\bibfnamefont {O.}~\bibnamefont {Ruchayskiy}},\ and\ \bibinfo {author} {\bibfnamefont {D.}~\bibnamefont {Iakubovskyi}},\ }\href {https://doi.org/10.1088/1475-7516/2009/03/005} {\bibfield  {journal} {\bibinfo  {journal} {J. Cosmol. Astropart. Phys.}\ }\textbf {\bibinfo {volume} {2009}}\bibinfo  {number} { (03)},\ \bibinfo {pages} {005}}\BibitemShut {NoStop}%
\bibitem [{\citenamefont {Di~Paolo}\ \emph {et~al.}(2018)\citenamefont {Di~Paolo}, \citenamefont {Nesti},\ and\ \citenamefont {Villante}}]{di_paolo_phase-space_2018}%
  \BibitemOpen
\bibfield  {number} {  }\bibfield  {author} {\bibinfo {author} {\bibfnamefont {C.}~\bibnamefont {Di~Paolo}}, \bibinfo {author} {\bibfnamefont {F.}~\bibnamefont {Nesti}},\ and\ \bibinfo {author} {\bibfnamefont {F.~L.}\ \bibnamefont {Villante}},\ }\href {https://doi.org/10.1093/mnras/sty091} {\bibfield  {journal} {\bibinfo  {journal} {Mon Not R Astron Soc}\ }\textbf {\bibinfo {volume} {475}},\ \bibinfo {pages} {5385} (\bibinfo {year} {2018})}\BibitemShut {NoStop}%
\bibitem [{\citenamefont {Alvey}\ \emph {et~al.}(2020)\citenamefont {Alvey}, \citenamefont {Sabti}, \citenamefont {Tiki}, \citenamefont {Blas}, \citenamefont {Bondarenko}, \citenamefont {Boyarsky}, \citenamefont {Escudero}, \citenamefont {Fairbairn}, \citenamefont {Orkney},\ and\ \citenamefont {Read}}]{alvey_new_2020}%
  \BibitemOpen
  \bibfield  {author} {\bibinfo {author} {\bibfnamefont {J.}~\bibnamefont {Alvey}}, \bibinfo {author} {\bibfnamefont {N.}~\bibnamefont {Sabti}}, \bibinfo {author} {\bibfnamefont {V.}~\bibnamefont {Tiki}}, \bibinfo {author} {\bibfnamefont {D.}~\bibnamefont {Blas}}, \bibinfo {author} {\bibfnamefont {K.}~\bibnamefont {Bondarenko}}, \bibinfo {author} {\bibfnamefont {A.}~\bibnamefont {Boyarsky}}, \bibinfo {author} {\bibfnamefont {M.}~\bibnamefont {Escudero}}, \bibinfo {author} {\bibfnamefont {M.}~\bibnamefont {Fairbairn}}, \bibinfo {author} {\bibfnamefont {M.}~\bibnamefont {Orkney}},\ and\ \bibinfo {author} {\bibfnamefont {J.~I.}\ \bibnamefont {Read}},\ }\href {https://doi.org/10.1093/mnras/staa3640} {\bibfield  {journal} {\bibinfo  {journal} {Monthly Notices of the Royal Astronomical Society}\ }\textbf {\bibinfo {volume} {501}},\ \bibinfo {pages} {1188} (\bibinfo {year} {2020})},\ \Eprint {https://arxiv.org/abs/2010.03572} {arXiv:2010.03572 [astro-ph]} \BibitemShut {NoStop}%
\bibitem [{\citenamefont {Stephenson}\ \emph {et~al.}(2012)\citenamefont {Stephenson}, \citenamefont {Goldman},\ and\ \citenamefont {Mckellar}}]{stephenson_jr_neutrino_2012}%
  \BibitemOpen
  \bibfield  {author} {\bibinfo {author} {\bibfnamefont {G.~J.}\ \bibnamefont {Stephenson}, \bibfnamefont {Jr.}}, \bibinfo {author} {\bibfnamefont {T.}~\bibnamefont {Goldman}},\ and\ \bibinfo {author} {\bibfnamefont {B.~H.~J.}\ \bibnamefont {Mckellar}},\ }\bibfield  {journal} {\bibinfo  {journal} {International Journal of Modern Physics A}\ }\href {https://doi.org/10.1142/S0217751X98001414} {10.1142/S0217751X98001414} (\bibinfo {year} {2012})\BibitemShut {NoStop}%
\bibitem [{\citenamefont {Smirnov}\ and\ \citenamefont {Xu}(2022)}]{smirnov_neutrino_2022}%
  \BibitemOpen
  \bibfield  {author} {\bibinfo {author} {\bibfnamefont {A.~Y.}\ \bibnamefont {Smirnov}}\ and\ \bibinfo {author} {\bibfnamefont {X.-J.}\ \bibnamefont {Xu}},\ }\href {https://doi.org/10.1007/JHEP08(2022)170} {\bibfield  {journal} {\bibinfo  {journal} {J. High Energ. Phys.}\ }\textbf {\bibinfo {volume} {2022}}\bibfield  {number} {\bibinfo  {number} { (8)},\ \bibinfo {pages} {170}},\ }\Eprint {https://arxiv.org/abs/2201.00939} {arXiv:2201.00939 [hep-ph]} \BibitemShut {NoStop}%
\bibitem [{\citenamefont {Kaplan}\ \emph {et~al.}(2025)\citenamefont {Kaplan}, \citenamefont {Luo},\ and\ \citenamefont {Rajendran}}]{kaplan_probing_2025}%
  \BibitemOpen
  \bibfield  {author} {\bibinfo {author} {\bibfnamefont {D.~E.}\ \bibnamefont {Kaplan}}, \bibinfo {author} {\bibfnamefont {X.}~\bibnamefont {Luo}},\ and\ \bibinfo {author} {\bibfnamefont {S.}~\bibnamefont {Rajendran}},\ }\href {https://doi.org/10.1103/PhysRevD.111.055019} {\bibfield  {journal} {\bibinfo  {journal} {Phys. Rev. D}\ }\textbf {\bibinfo {volume} {111}},\ \bibinfo {pages} {055019} (\bibinfo {year} {2025})}\BibitemShut {NoStop}%
\bibitem [{\citenamefont {Burkert}(2015)}]{burkert_structure_2015}%
  \BibitemOpen
  \bibfield  {author} {\bibinfo {author} {\bibfnamefont {A.}~\bibnamefont {Burkert}},\ }\href {https://doi.org/10.1088/0004-637X/808/2/158} {\bibfield  {journal} {\bibinfo  {journal} {ApJ}\ }\textbf {\bibinfo {volume} {808}},\ \bibinfo {pages} {158} (\bibinfo {year} {2015})}\BibitemShut {NoStop}%
\bibitem [{\citenamefont {Kormendy}\ and\ \citenamefont {Freeman}(2016)}]{kormendy_scaling_2016}%
  \BibitemOpen
  \bibfield  {author} {\bibinfo {author} {\bibfnamefont {J.}~\bibnamefont {Kormendy}}\ and\ \bibinfo {author} {\bibfnamefont {K.~C.}\ \bibnamefont {Freeman}},\ }\href {https://doi.org/10.3847/0004-637X/817/2/84} {\bibfield  {journal} {\bibinfo  {journal} {ApJ}\ }\textbf {\bibinfo {volume} {817}},\ \bibinfo {pages} {84} (\bibinfo {year} {2016})}\BibitemShut {NoStop}%
\bibitem [{\citenamefont {S{\'a}nchez~Almeida}(2025)}]{sanchez_almeida_implications_2025}%
  \BibitemOpen
  \bibfield  {author} {\bibinfo {author} {\bibfnamefont {J.}~\bibnamefont {S{\'a}nchez~Almeida}},\ }\href {https://doi.org/10.3390/galaxies13010006} {\bibfield  {journal} {\bibinfo  {journal} {Galaxies}\ }\textbf {\bibinfo {volume} {13}},\ \bibinfo {pages} {6} (\bibinfo {year} {2025})}\BibitemShut {NoStop}%
\bibitem [{\citenamefont {Chavanis}(2020)}]{chavanis_statistical_2020}%
  \BibitemOpen
  \bibfield  {author} {\bibinfo {author} {\bibfnamefont {P.-H.}\ \bibnamefont {Chavanis}},\ }\href {https://doi.org/10.1140/epjp/s13360-020-00268-0} {\bibfield  {journal} {\bibinfo  {journal} {Eur. Phys. J. Plus}\ }\textbf {\bibinfo {volume} {135}},\ \bibinfo {pages} {290} (\bibinfo {year} {2020})}\BibitemShut {NoStop}%
\bibitem [{\citenamefont {Chavanis}(2022)}]{chavanis_predictive_2022}%
  \BibitemOpen
  \bibfield  {author} {\bibinfo {author} {\bibfnamefont {P.-H.}\ \bibnamefont {Chavanis}},\ }\href {https://doi.org/10.1103/PhysRevD.106.043538} {\bibfield  {journal} {\bibinfo  {journal} {Phys. Rev. D}\ }\textbf {\bibinfo {volume} {106}},\ \bibinfo {pages} {043538} (\bibinfo {year} {2022})}\BibitemShut {NoStop}%
\bibitem [{\citenamefont {Arg{\"u}elles}\ \emph {et~al.}(2018)\citenamefont {Arg{\"u}elles}, \citenamefont {Krut}, \citenamefont {Rueda},\ and\ \citenamefont {Ruffini}}]{arguelles_novel_2018}%
  \BibitemOpen
  \bibfield  {author} {\bibinfo {author} {\bibfnamefont {C.~R.}\ \bibnamefont {Arg{\"u}elles}}, \bibinfo {author} {\bibfnamefont {A.}~\bibnamefont {Krut}}, \bibinfo {author} {\bibfnamefont {J.~A.}\ \bibnamefont {Rueda}},\ and\ \bibinfo {author} {\bibfnamefont {R.}~\bibnamefont {Ruffini}},\ }\href {https://doi.org/10.1016/j.dark.2018.07.002} {\bibfield  {journal} {\bibinfo  {journal} {Physics of the Dark Universe}\ }\textbf {\bibinfo {volume} {21}},\ \bibinfo {pages} {82} (\bibinfo {year} {2018})}\BibitemShut {NoStop}%
\bibitem [{\citenamefont {Arg{\"u}elles}\ \emph {et~al.}(2019)\citenamefont {Arg{\"u}elles}, \citenamefont {Krut}, \citenamefont {Rueda},\ and\ \citenamefont {Ruffini}}]{arguelles_novel_2019}%
  \BibitemOpen
  \bibfield  {author} {\bibinfo {author} {\bibfnamefont {C.~R.}\ \bibnamefont {Arg{\"u}elles}}, \bibinfo {author} {\bibfnamefont {A.}~\bibnamefont {Krut}}, \bibinfo {author} {\bibfnamefont {J.~A.}\ \bibnamefont {Rueda}},\ and\ \bibinfo {author} {\bibfnamefont {R.}~\bibnamefont {Ruffini}},\ }\href {https://doi.org/10.1016/j.dark.2019.100278} {\bibfield  {journal} {\bibinfo  {journal} {Physics of the Dark Universe}\ }\textbf {\bibinfo {volume} {24}},\ \bibinfo {pages} {100278} (\bibinfo {year} {2019})}\BibitemShut {NoStop}%
\bibitem [{\citenamefont {Arg{\"u}elles}\ \emph {et~al.}(2021)\citenamefont {Arg{\"u}elles}, \citenamefont {D{\'i}az}, \citenamefont {Krut},\ and\ \citenamefont {Yunis}}]{arguelles_formation_2021}%
  \BibitemOpen
  \bibfield  {author} {\bibinfo {author} {\bibfnamefont {C.~R.}\ \bibnamefont {Arg{\"u}elles}}, \bibinfo {author} {\bibfnamefont {M.~I.}\ \bibnamefont {D{\'i}az}}, \bibinfo {author} {\bibfnamefont {A.}~\bibnamefont {Krut}},\ and\ \bibinfo {author} {\bibfnamefont {R.}~\bibnamefont {Yunis}},\ }\href {https://doi.org/10.1093/mnras/staa3986} {\bibfield  {journal} {\bibinfo  {journal} {Monthly Notices of the Royal Astronomical Society}\ }\textbf {\bibinfo {volume} {502}},\ \bibinfo {pages} {4227} (\bibinfo {year} {2021})}\BibitemShut {NoStop}%
\bibitem [{\citenamefont {Dom{\`e}nech}\ \emph {et~al.}(2023)\citenamefont {Dom{\`e}nech}, \citenamefont {Inman}, \citenamefont {Kusenko},\ and\ \citenamefont {Sasaki}}]{domenech_halo_2023}%
  \BibitemOpen
  \bibfield  {author} {\bibinfo {author} {\bibfnamefont {G.}~\bibnamefont {Dom{\`e}nech}}, \bibinfo {author} {\bibfnamefont {D.}~\bibnamefont {Inman}}, \bibinfo {author} {\bibfnamefont {A.}~\bibnamefont {Kusenko}},\ and\ \bibinfo {author} {\bibfnamefont {M.}~\bibnamefont {Sasaki}},\ }\href {https://doi.org/10.1103/PhysRevD.108.103543} {\bibfield  {journal} {\bibinfo  {journal} {Phys. Rev. D}\ }\textbf {\bibinfo {volume} {108}},\ \bibinfo {pages} {103543} (\bibinfo {year} {2023})}\BibitemShut {NoStop}%
\bibitem [{\citenamefont {{Lynden-Bell}}(1967)}]{lynden-bell_statistical_1967}%
  \BibitemOpen
  \bibfield  {author} {\bibinfo {author} {\bibfnamefont {D.}~\bibnamefont {{Lynden-Bell}}},\ }\href {https://doi.org/10.1093/mnras/136.1.101} {\bibfield  {journal} {\bibinfo  {journal} {Monthly Notices of the Royal Astronomical Society}\ }\textbf {\bibinfo {volume} {136}},\ \bibinfo {pages} {101} (\bibinfo {year} {1967})}\BibitemShut {NoStop}%
\bibitem [{\citenamefont {Chavanis}\ and\ \citenamefont {Sommeria}(1998)}]{chavanis_degenerate_1998}%
  \BibitemOpen
  \bibfield  {author} {\bibinfo {author} {\bibfnamefont {P.-H.}\ \bibnamefont {Chavanis}}\ and\ \bibinfo {author} {\bibfnamefont {J.}~\bibnamefont {Sommeria}},\ }\href {https://doi.org/10.1046/j.1365-8711.1998.01414.x} {\bibfield  {journal} {\bibinfo  {journal} {Mon Not R Astron Soc}\ }\textbf {\bibinfo {volume} {296}},\ \bibinfo {pages} {569} (\bibinfo {year} {1998})}\BibitemShut {NoStop}%
\bibitem [{\citenamefont {Barkana}\ and\ \citenamefont {Loeb}(2001)}]{barkana_beginning_2001}%
  \BibitemOpen
  \bibfield  {author} {\bibinfo {author} {\bibfnamefont {R.}~\bibnamefont {Barkana}}\ and\ \bibinfo {author} {\bibfnamefont {A.}~\bibnamefont {Loeb}},\ }\href {https://doi.org/10.1016/S0370-1573(01)00019-9} {\bibfield  {journal} {\bibinfo  {journal} {Physics Reports}\ }\textbf {\bibinfo {volume} {349}},\ \bibinfo {pages} {125} (\bibinfo {year} {2001})}\BibitemShut {NoStop}%
\bibitem [{\citenamefont {Bode}\ \emph {et~al.}(2001)\citenamefont {Bode}, \citenamefont {Ostriker},\ and\ \citenamefont {Turok}}]{bode_halo_2001}%
  \BibitemOpen
  \bibfield  {author} {\bibinfo {author} {\bibfnamefont {P.}~\bibnamefont {Bode}}, \bibinfo {author} {\bibfnamefont {J.~P.}\ \bibnamefont {Ostriker}},\ and\ \bibinfo {author} {\bibfnamefont {N.}~\bibnamefont {Turok}},\ }\href {https://doi.org/10.1086/321541} {\bibfield  {journal} {\bibinfo  {journal} {ApJ}\ }\textbf {\bibinfo {volume} {556}},\ \bibinfo {pages} {93} (\bibinfo {year} {2001})}\BibitemShut {NoStop}%
\bibitem [{\citenamefont {Lesgourgues}\ and\ \citenamefont {Pastor}(2006)}]{lesgourgues_massive_2006}%
  \BibitemOpen
  \bibfield  {author} {\bibinfo {author} {\bibfnamefont {J.}~\bibnamefont {Lesgourgues}}\ and\ \bibinfo {author} {\bibfnamefont {S.}~\bibnamefont {Pastor}},\ }\href {https://doi.org/10.1016/j.physrep.2006.04.001} {\bibfield  {journal} {\bibinfo  {journal} {Physics Reports}\ }\textbf {\bibinfo {volume} {429}},\ \bibinfo {pages} {307} (\bibinfo {year} {2006})}\BibitemShut {NoStop}%
\bibitem [{\citenamefont {Markevitch}\ \emph {et~al.}(2004)\citenamefont {Markevitch}, \citenamefont {Gonzalez}, \citenamefont {Clowe}, \citenamefont {Vikhlinin}, \citenamefont {Forman}, \citenamefont {Jones}, \citenamefont {Murray},\ and\ \citenamefont {Tucker}}]{markevitch_direct_2004}%
  \BibitemOpen
  \bibfield  {author} {\bibinfo {author} {\bibfnamefont {M.}~\bibnamefont {Markevitch}}, \bibinfo {author} {\bibfnamefont {A.~H.}\ \bibnamefont {Gonzalez}}, \bibinfo {author} {\bibfnamefont {D.}~\bibnamefont {Clowe}}, \bibinfo {author} {\bibfnamefont {A.}~\bibnamefont {Vikhlinin}}, \bibinfo {author} {\bibfnamefont {W.}~\bibnamefont {Forman}}, \bibinfo {author} {\bibfnamefont {C.}~\bibnamefont {Jones}}, \bibinfo {author} {\bibfnamefont {S.}~\bibnamefont {Murray}},\ and\ \bibinfo {author} {\bibfnamefont {W.}~\bibnamefont {Tucker}},\ }\href {https://doi.org/10.1086/383178} {\bibfield  {journal} {\bibinfo  {journal} {ApJ}\ }\textbf {\bibinfo {volume} {606}},\ \bibinfo {pages} {819} (\bibinfo {year} {2004})}\BibitemShut {NoStop}%
\bibitem [{\citenamefont {Randall}\ \emph {et~al.}(2008)\citenamefont {Randall}, \citenamefont {Markevitch}, \citenamefont {Clowe}, \citenamefont {Gonzalez},\ and\ \citenamefont {Brada{\v c}}}]{randall_constraints_2008}%
  \BibitemOpen
  \bibfield  {author} {\bibinfo {author} {\bibfnamefont {S.~W.}\ \bibnamefont {Randall}}, \bibinfo {author} {\bibfnamefont {M.}~\bibnamefont {Markevitch}}, \bibinfo {author} {\bibfnamefont {D.}~\bibnamefont {Clowe}}, \bibinfo {author} {\bibfnamefont {A.~H.}\ \bibnamefont {Gonzalez}},\ and\ \bibinfo {author} {\bibfnamefont {M.}~\bibnamefont {Brada{\v c}}},\ }\href {https://doi.org/10.1086/587859} {\bibfield  {journal} {\bibinfo  {journal} {ApJ}\ }\textbf {\bibinfo {volume} {679}},\ \bibinfo {pages} {1173} (\bibinfo {year} {2008})}\BibitemShut {NoStop}%
\bibitem [{\citenamefont {Bogorad}\ \emph {et~al.}(2025)\citenamefont {Bogorad}, \citenamefont {Graham},\ and\ \citenamefont {Ramani}}]{bogorad_coherent_2025}%
  \BibitemOpen
  \bibfield  {author} {\bibinfo {author} {\bibfnamefont {Z.}~\bibnamefont {Bogorad}}, \bibinfo {author} {\bibfnamefont {P.~W.}\ \bibnamefont {Graham}},\ and\ \bibinfo {author} {\bibfnamefont {H.}~\bibnamefont {Ramani}},\ }\href {https://doi.org/10.1088/1475-7516/2025/03/067} {\bibfield  {journal} {\bibinfo  {journal} {J. Cosmol. Astropart. Phys.}\ }\textbf {\bibinfo {volume} {2025}}\bibinfo  {number} { (03)},\ \bibinfo {pages} {067}}\BibitemShut {NoStop}%
\bibitem [{\citenamefont {Bottaro}\ \emph {et~al.}(2024)\citenamefont {Bottaro}, \citenamefont {Castorina}, \citenamefont {Costa}, \citenamefont {Redigolo},\ and\ \citenamefont {Salvioni}}]{bottaro_unveiling_2024}%
  \BibitemOpen
\bibfield  {number} {  }\bibfield  {author} {\bibinfo {author} {\bibfnamefont {S.}~\bibnamefont {Bottaro}}, \bibinfo {author} {\bibfnamefont {E.}~\bibnamefont {Castorina}}, \bibinfo {author} {\bibfnamefont {M.}~\bibnamefont {Costa}}, \bibinfo {author} {\bibfnamefont {D.}~\bibnamefont {Redigolo}},\ and\ \bibinfo {author} {\bibfnamefont {E.}~\bibnamefont {Salvioni}},\ }\href {https://doi.org/10.1103/PhysRevLett.132.201002} {\bibfield  {journal} {\bibinfo  {journal} {Phys. Rev. Lett.}\ }\textbf {\bibinfo {volume} {132}},\ \bibinfo {pages} {201002} (\bibinfo {year} {2024})}\BibitemShut {NoStop}%
\bibitem [{\citenamefont {Bottaro}\ \emph {et~al.}(2025)\citenamefont {Bottaro}, \citenamefont {Castorina}, \citenamefont {Costa}, \citenamefont {Redigolo},\ and\ \citenamefont {Salvioni}}]{bottaro_100_2025}%
  \BibitemOpen
  \bibfield  {author} {\bibinfo {author} {\bibfnamefont {S.}~\bibnamefont {Bottaro}}, \bibinfo {author} {\bibfnamefont {E.}~\bibnamefont {Castorina}}, \bibinfo {author} {\bibfnamefont {M.}~\bibnamefont {Costa}}, \bibinfo {author} {\bibfnamefont {D.}~\bibnamefont {Redigolo}},\ and\ \bibinfo {author} {\bibfnamefont {E.}~\bibnamefont {Salvioni}},\ }\href {https://doi.org/10.1103/gc78-96l5} {\bibfield  {journal} {\bibinfo  {journal} {Phys. Rev. D}\ }\textbf {\bibinfo {volume} {112}},\ \bibinfo {pages} {023525} (\bibinfo {year} {2025})}\BibitemShut {NoStop}%
\bibitem [{\citenamefont {Kesden}\ and\ \citenamefont {Kamionkowski}(2006)}]{kesden_tidal_2006}%
  \BibitemOpen
  \bibfield  {author} {\bibinfo {author} {\bibfnamefont {M.}~\bibnamefont {Kesden}}\ and\ \bibinfo {author} {\bibfnamefont {M.}~\bibnamefont {Kamionkowski}},\ }\href {https://doi.org/10.1103/PhysRevD.74.083007} {\bibfield  {journal} {\bibinfo  {journal} {Phys. Rev. D}\ }\textbf {\bibinfo {volume} {74}},\ \bibinfo {pages} {083007} (\bibinfo {year} {2006})}\BibitemShut {NoStop}%
\bibitem [{\citenamefont {Yeh}\ \emph {et~al.}(2026)\citenamefont {Yeh}, \citenamefont {Olive}, \citenamefont {Fields}, \citenamefont {Aver}, \citenamefont {Pogge}, \citenamefont {Rogers}, \citenamefont {Skillman},\ and\ \citenamefont {Weller}}]{yeh_lbt_2026}%
  \BibitemOpen
  \bibfield  {author} {\bibinfo {author} {\bibfnamefont {T.-H.}\ \bibnamefont {Yeh}}, \bibinfo {author} {\bibfnamefont {K.~A.}\ \bibnamefont {Olive}}, \bibinfo {author} {\bibfnamefont {B.~D.}\ \bibnamefont {Fields}}, \bibinfo {author} {\bibfnamefont {E.}~\bibnamefont {Aver}}, \bibinfo {author} {\bibfnamefont {R.~W.}\ \bibnamefont {Pogge}}, \bibinfo {author} {\bibfnamefont {N.~S.~J.}\ \bibnamefont {Rogers}}, \bibinfo {author} {\bibfnamefont {E.~D.}\ \bibnamefont {Skillman}},\ and\ \bibinfo {author} {\bibfnamefont {M.~K.}\ \bibnamefont {Weller}},\ }\href {https://doi.org/10.48550/arXiv.2601.22239} {\bibinfo {title} {The {{LBT}} \${{Y}}\_\textbraceleft\textbackslash rm p\textbraceright\$ {{Project V}}: {{Cosmological Implications}} of a {{New Determination}} of {{Primordial}} \$\textasciicircum 4\${{He}}}} (\bibinfo {year} {2026}),\ \Eprint {https://arxiv.org/abs/2601.22239} {arXiv:2601.22239 [astro-ph.CO]} \BibitemShut {NoStop}%
\bibitem [{\citenamefont {Walecka}(1974)}]{walecka_theory_1974}%
  \BibitemOpen
  \bibfield  {author} {\bibinfo {author} {\bibfnamefont {J.}~\bibnamefont {Walecka}},\ }\href {https://doi.org/10.1016/0003-4916(74)90208-5} {\bibfield  {journal} {\bibinfo  {journal} {Annals of Physics}\ }\textbf {\bibinfo {volume} {83}},\ \bibinfo {pages} {491} (\bibinfo {year} {1974})}\BibitemShut {NoStop}%
\bibitem [{\citenamefont {{DES Collaboration}}\ \emph {et~al.}(2021)\citenamefont {{DES Collaboration}}, \citenamefont {Nadler}, \citenamefont {{Drlica-Wagner}}, \citenamefont {Bechtol}, \citenamefont {Mau}, \citenamefont {Wechsler}, \citenamefont {Gluscevic}, \citenamefont {Boddy}, \citenamefont {Pace}, \citenamefont {Li}, \citenamefont {McNanna}, \citenamefont {Riley}, \citenamefont {{Garc{\'i}a-Bellido}}, \citenamefont {Mao}, \citenamefont {Green}, \citenamefont {Burke}, \citenamefont {Peter}, \citenamefont {Jain}, \citenamefont {Abbott}, \citenamefont {Aguena}, \citenamefont {Allam}, \citenamefont {Annis}, \citenamefont {Avila}, \citenamefont {Brooks}, \citenamefont {Carrasco~Kind}, \citenamefont {Carretero}, \citenamefont {Costanzi}, \citenamefont {{da Costa}}, \citenamefont {De~Vicente}, \citenamefont {Desai}, \citenamefont {Diehl}, \citenamefont {Doel}, \citenamefont {Everett}, \citenamefont {Evrard}, \citenamefont {Flaugher}, \citenamefont {Frieman}, \citenamefont {Gerdes}, \citenamefont {Gruen},
  \citenamefont {Gruendl}, \citenamefont {Gschwend}, \citenamefont {Gutierrez}, \citenamefont {Hinton}, \citenamefont {Honscheid}, \citenamefont {Huterer}, \citenamefont {James}, \citenamefont {Krause}, \citenamefont {Kuehn}, \citenamefont {Kuropatkin}, \citenamefont {Lahav}, \citenamefont {Maia}, \citenamefont {Marshall}, \citenamefont {Menanteau}, \citenamefont {Miquel}, \citenamefont {Palmese}, \citenamefont {{Paz-Chinch{\'o}n}}, \citenamefont {Plazas}, \citenamefont {Romer}, \citenamefont {Sanchez}, \citenamefont {Scarpine}, \citenamefont {Serrano}, \citenamefont {{Sevilla-Noarbe}}, \citenamefont {Smith}, \citenamefont {{Soares-Santos}}, \citenamefont {Suchyta}, \citenamefont {Swanson}, \citenamefont {Tarle}, \citenamefont {Tucker}, \citenamefont {Walker},\ and\ \citenamefont {Wester}}]{des_collaboration_constraints_2021}%
  \BibitemOpen
  \bibfield  {author} {\bibinfo {author} {\bibnamefont {{DES Collaboration}}}, \bibinfo {author} {\bibfnamefont {E.~O.}\ \bibnamefont {Nadler}}, \bibinfo {author} {\bibfnamefont {A.}~\bibnamefont {{Drlica-Wagner}}}, \bibinfo {author} {\bibfnamefont {K.}~\bibnamefont {Bechtol}}, \bibinfo {author} {\bibfnamefont {S.}~\bibnamefont {Mau}}, \bibinfo {author} {\bibfnamefont {R.~H.}\ \bibnamefont {Wechsler}}, \bibinfo {author} {\bibfnamefont {V.}~\bibnamefont {Gluscevic}}, \bibinfo {author} {\bibfnamefont {K.}~\bibnamefont {Boddy}}, \bibinfo {author} {\bibfnamefont {A.~B.}\ \bibnamefont {Pace}}, \bibinfo {author} {\bibfnamefont {T.~S.}\ \bibnamefont {Li}}, \bibinfo {author} {\bibfnamefont {M.}~\bibnamefont {McNanna}}, \bibinfo {author} {\bibfnamefont {A.~H.}\ \bibnamefont {Riley}}, \bibinfo {author} {\bibfnamefont {J.}~\bibnamefont {{Garc{\'i}a-Bellido}}}, \bibinfo {author} {\bibfnamefont {Y.-Y.}\ \bibnamefont {Mao}}, \bibinfo {author} {\bibfnamefont {G.}~\bibnamefont {Green}}, \bibinfo {author} {\bibfnamefont
  {D.~L.}\ \bibnamefont {Burke}}, \bibinfo {author} {\bibfnamefont {A.}~\bibnamefont {Peter}}, \bibinfo {author} {\bibfnamefont {B.}~\bibnamefont {Jain}}, \bibinfo {author} {\bibfnamefont {T.~M.~C.}\ \bibnamefont {Abbott}}, \bibinfo {author} {\bibfnamefont {M.}~\bibnamefont {Aguena}}, \bibinfo {author} {\bibfnamefont {S.}~\bibnamefont {Allam}}, \bibinfo {author} {\bibfnamefont {J.}~\bibnamefont {Annis}}, \bibinfo {author} {\bibfnamefont {S.}~\bibnamefont {Avila}}, \bibinfo {author} {\bibfnamefont {D.}~\bibnamefont {Brooks}}, \bibinfo {author} {\bibfnamefont {M.}~\bibnamefont {Carrasco~Kind}}, \bibinfo {author} {\bibfnamefont {J.}~\bibnamefont {Carretero}}, \bibinfo {author} {\bibfnamefont {M.}~\bibnamefont {Costanzi}}, \bibinfo {author} {\bibfnamefont {L.~N.}\ \bibnamefont {{da Costa}}}, \bibinfo {author} {\bibfnamefont {J.}~\bibnamefont {De~Vicente}}, \bibinfo {author} {\bibfnamefont {S.}~\bibnamefont {Desai}}, \bibinfo {author} {\bibfnamefont {H.~T.}\ \bibnamefont {Diehl}}, \bibinfo {author} {\bibfnamefont
  {P.}~\bibnamefont {Doel}}, \bibinfo {author} {\bibfnamefont {S.}~\bibnamefont {Everett}}, \bibinfo {author} {\bibfnamefont {A.~E.}\ \bibnamefont {Evrard}}, \bibinfo {author} {\bibfnamefont {B.}~\bibnamefont {Flaugher}}, \bibinfo {author} {\bibfnamefont {J.}~\bibnamefont {Frieman}}, \bibinfo {author} {\bibfnamefont {D.~W.}\ \bibnamefont {Gerdes}}, \bibinfo {author} {\bibfnamefont {D.}~\bibnamefont {Gruen}}, \bibinfo {author} {\bibfnamefont {R.~A.}\ \bibnamefont {Gruendl}}, \bibinfo {author} {\bibfnamefont {J.}~\bibnamefont {Gschwend}}, \bibinfo {author} {\bibfnamefont {G.}~\bibnamefont {Gutierrez}}, \bibinfo {author} {\bibfnamefont {S.~R.}\ \bibnamefont {Hinton}}, \bibinfo {author} {\bibfnamefont {K.}~\bibnamefont {Honscheid}}, \bibinfo {author} {\bibfnamefont {D.}~\bibnamefont {Huterer}}, \bibinfo {author} {\bibfnamefont {D.~J.}\ \bibnamefont {James}}, \bibinfo {author} {\bibfnamefont {E.}~\bibnamefont {Krause}}, \bibinfo {author} {\bibfnamefont {K.}~\bibnamefont {Kuehn}}, \bibinfo {author} {\bibfnamefont
  {N.}~\bibnamefont {Kuropatkin}}, \bibinfo {author} {\bibfnamefont {O.}~\bibnamefont {Lahav}}, \bibinfo {author} {\bibfnamefont {M.~A.~G.}\ \bibnamefont {Maia}}, \bibinfo {author} {\bibfnamefont {J.~L.}\ \bibnamefont {Marshall}}, \bibinfo {author} {\bibfnamefont {F.}~\bibnamefont {Menanteau}}, \bibinfo {author} {\bibfnamefont {R.}~\bibnamefont {Miquel}}, \bibinfo {author} {\bibfnamefont {A.}~\bibnamefont {Palmese}}, \bibinfo {author} {\bibfnamefont {F.}~\bibnamefont {{Paz-Chinch{\'o}n}}}, \bibinfo {author} {\bibfnamefont {A.~A.}\ \bibnamefont {Plazas}}, \bibinfo {author} {\bibfnamefont {A.~K.}\ \bibnamefont {Romer}}, \bibinfo {author} {\bibfnamefont {E.}~\bibnamefont {Sanchez}}, \bibinfo {author} {\bibfnamefont {V.}~\bibnamefont {Scarpine}}, \bibinfo {author} {\bibfnamefont {S.}~\bibnamefont {Serrano}}, \bibinfo {author} {\bibfnamefont {I.}~\bibnamefont {{Sevilla-Noarbe}}}, \bibinfo {author} {\bibfnamefont {M.}~\bibnamefont {Smith}}, \bibinfo {author} {\bibfnamefont {M.}~\bibnamefont {{Soares-Santos}}},
  \bibinfo {author} {\bibfnamefont {E.}~\bibnamefont {Suchyta}}, \bibinfo {author} {\bibfnamefont {M.~E.~C.}\ \bibnamefont {Swanson}}, \bibinfo {author} {\bibfnamefont {G.}~\bibnamefont {Tarle}}, \bibinfo {author} {\bibfnamefont {D.~L.}\ \bibnamefont {Tucker}}, \bibinfo {author} {\bibfnamefont {A.~R.}\ \bibnamefont {Walker}},\ and\ \bibinfo {author} {\bibfnamefont {W.}~\bibnamefont {Wester}},\ }\href {https://doi.org/10.1103/PhysRevLett.126.091101} {\bibfield  {journal} {\bibinfo  {journal} {Phys. Rev. Lett.}\ }\textbf {\bibinfo {volume} {126}},\ \bibinfo {pages} {091101} (\bibinfo {year} {2021})}\BibitemShut {NoStop}%
\bibitem [{\citenamefont {Afshordi}\ \emph {et~al.}(2003)\citenamefont {Afshordi}, \citenamefont {McDonald},\ and\ \citenamefont {Spergel}}]{afshordi_primordial_2003}%
  \BibitemOpen
  \bibfield  {author} {\bibinfo {author} {\bibfnamefont {N.}~\bibnamefont {Afshordi}}, \bibinfo {author} {\bibfnamefont {P.}~\bibnamefont {McDonald}},\ and\ \bibinfo {author} {\bibfnamefont {D.~N.}\ \bibnamefont {Spergel}},\ }\href {https://doi.org/10.1086/378763} {\bibfield  {journal} {\bibinfo  {journal} {ApJ}\ }\textbf {\bibinfo {volume} {594}},\ \bibinfo {pages} {L71} (\bibinfo {year} {2003})}\BibitemShut {NoStop}%
\bibitem [{\citenamefont {Peebles}(1980)}]{peebles_large-scale_1980}%
  \BibitemOpen
  \bibfield  {author} {\bibinfo {author} {\bibfnamefont {P.~J.~E.}\ \bibnamefont {Peebles}},\ }\href@noop {} {\emph {\bibinfo {title} {The Large-Scale Structure of the Universe}}}\ (\bibinfo  {publisher} {Princeton University Press},\ \bibinfo {year} {1980})\BibitemShut {NoStop}%
\bibitem [{\citenamefont {Murgia}\ \emph {et~al.}(2017)\citenamefont {Murgia}, \citenamefont {Merle}, \citenamefont {Viel}, \citenamefont {Totzauer},\ and\ \citenamefont {Schneider}}]{murgia_non-cold_2017}%
  \BibitemOpen
  \bibfield  {author} {\bibinfo {author} {\bibfnamefont {R.}~\bibnamefont {Murgia}}, \bibinfo {author} {\bibfnamefont {A.}~\bibnamefont {Merle}}, \bibinfo {author} {\bibfnamefont {M.}~\bibnamefont {Viel}}, \bibinfo {author} {\bibfnamefont {M.}~\bibnamefont {Totzauer}},\ and\ \bibinfo {author} {\bibfnamefont {A.}~\bibnamefont {Schneider}},\ }\href {https://doi.org/10.1088/1475-7516/2017/11/046} {\bibfield  {journal} {\bibinfo  {journal} {J. Cosmol. Astropart. Phys.}\ }\textbf {\bibinfo {volume} {2017}}\bibinfo  {number} { (11)},\ \bibinfo {pages} {046}}\BibitemShut {NoStop}%
\bibitem [{\citenamefont {Graham}\ and\ \citenamefont {Ramani}(2024)}]{graham_constraints_2024}%
  \BibitemOpen
\bibfield  {number} {  }\bibfield  {author} {\bibinfo {author} {\bibfnamefont {P.~W.}\ \bibnamefont {Graham}}\ and\ \bibinfo {author} {\bibfnamefont {H.}~\bibnamefont {Ramani}},\ }\href {https://doi.org/10.1103/PhysRevD.110.075012} {\bibfield  {journal} {\bibinfo  {journal} {Phys. Rev. D}\ }\textbf {\bibinfo {volume} {110}},\ \bibinfo {pages} {075012} (\bibinfo {year} {2024})}\BibitemShut {NoStop}%
\bibitem [{\citenamefont {Gasser}\ and\ \citenamefont {Leutwyler}(1984)}]{gasser_chiral_1984}%
  \BibitemOpen
  \bibfield  {author} {\bibinfo {author} {\bibfnamefont {J.}~\bibnamefont {Gasser}}\ and\ \bibinfo {author} {\bibfnamefont {H.}~\bibnamefont {Leutwyler}},\ }\href {https://doi.org/10.1016/0003-4916(84)90242-2} {\bibfield  {journal} {\bibinfo  {journal} {Annals of Physics}\ }\textbf {\bibinfo {volume} {158}},\ \bibinfo {pages} {142} (\bibinfo {year} {1984})}\BibitemShut {NoStop}%
\bibitem [{\citenamefont {Panico}\ and\ \citenamefont {Wulzer}(2016)}]{panico_composite_2016}%
  \BibitemOpen
  \bibfield  {author} {\bibinfo {author} {\bibfnamefont {G.}~\bibnamefont {Panico}}\ and\ \bibinfo {author} {\bibfnamefont {A.}~\bibnamefont {Wulzer}},\ }\href {https://doi.org/10.1007/978-3-319-22617-0} {\emph {\bibinfo {title} {The {{Composite Nambu-Goldstone Higgs}}}}},\ \bibinfo {series} {Lecture {{Notes}} in {{Physics}}}, Vol.\ \bibinfo {volume} {913}\ (\bibinfo  {publisher} {Springer International Publishing},\ \bibinfo {address} {Cham},\ \bibinfo {year} {2016})\BibitemShut {NoStop}%
\bibitem [{\citenamefont {Schmaltz}\ and\ \citenamefont {{Tucker-Smith}}(2005)}]{schmaltz_little_2005}%
  \BibitemOpen
  \bibfield  {author} {\bibinfo {author} {\bibfnamefont {M.}~\bibnamefont {Schmaltz}}\ and\ \bibinfo {author} {\bibfnamefont {D.}~\bibnamefont {{Tucker-Smith}}},\ }\href {https://doi.org/10.1146/annurev.nucl.55.090704.151502} {\bibfield  {journal} {\bibinfo  {journal} {Annual Review of Nuclear and Particle Science}\ }\textbf {\bibinfo {volume} {55}},\ \bibinfo {pages} {229} (\bibinfo {year} {2005})}\BibitemShut {NoStop}%
\bibitem [{\citenamefont {Pacucci}\ \emph {et~al.}(2023)\citenamefont {Pacucci}, \citenamefont {Nguyen}, \citenamefont {Carniani}, \citenamefont {Maiolino},\ and\ \citenamefont {Fan}}]{pacucci_jwst_2023}%
  \BibitemOpen
  \bibfield  {author} {\bibinfo {author} {\bibfnamefont {F.}~\bibnamefont {Pacucci}}, \bibinfo {author} {\bibfnamefont {B.}~\bibnamefont {Nguyen}}, \bibinfo {author} {\bibfnamefont {S.}~\bibnamefont {Carniani}}, \bibinfo {author} {\bibfnamefont {R.}~\bibnamefont {Maiolino}},\ and\ \bibinfo {author} {\bibfnamefont {X.}~\bibnamefont {Fan}},\ }\href {https://doi.org/10.3847/2041-8213/ad0158} {\bibfield  {journal} {\bibinfo  {journal} {ApJL}\ }\textbf {\bibinfo {volume} {957}},\ \bibinfo {pages} {L3} (\bibinfo {year} {2023})}\BibitemShut {NoStop}%
\bibitem [{\citenamefont {Matthee}\ \emph {et~al.}(2024)\citenamefont {Matthee}, \citenamefont {Naidu}, \citenamefont {Brammer}, \citenamefont {Chisholm}, \citenamefont {Eilers}, \citenamefont {Goulding}, \citenamefont {Greene}, \citenamefont {Kashino}, \citenamefont {Labbe}, \citenamefont {Lilly}, \citenamefont {Mackenzie}, \citenamefont {Oesch}, \citenamefont {Weibel}, \citenamefont {Wuyts}, \citenamefont {Xiao}, \citenamefont {Bordoloi}, \citenamefont {Bouwens}, \citenamefont {{van Dokkum}}, \citenamefont {Illingworth}, \citenamefont {Kramarenko}, \citenamefont {Maseda}, \citenamefont {Mason}, \citenamefont {Meyer}, \citenamefont {Nelson}, \citenamefont {Reddy}, \citenamefont {Shivaei}, \citenamefont {Simcoe},\ and\ \citenamefont {Yue}}]{matthee_little_2024}%
  \BibitemOpen
  \bibfield  {author} {\bibinfo {author} {\bibfnamefont {J.}~\bibnamefont {Matthee}}, \bibinfo {author} {\bibfnamefont {R.~P.}\ \bibnamefont {Naidu}}, \bibinfo {author} {\bibfnamefont {G.}~\bibnamefont {Brammer}}, \bibinfo {author} {\bibfnamefont {J.}~\bibnamefont {Chisholm}}, \bibinfo {author} {\bibfnamefont {A.-C.}\ \bibnamefont {Eilers}}, \bibinfo {author} {\bibfnamefont {A.}~\bibnamefont {Goulding}}, \bibinfo {author} {\bibfnamefont {J.}~\bibnamefont {Greene}}, \bibinfo {author} {\bibfnamefont {D.}~\bibnamefont {Kashino}}, \bibinfo {author} {\bibfnamefont {I.}~\bibnamefont {Labbe}}, \bibinfo {author} {\bibfnamefont {S.~J.}\ \bibnamefont {Lilly}}, \bibinfo {author} {\bibfnamefont {R.}~\bibnamefont {Mackenzie}}, \bibinfo {author} {\bibfnamefont {P.~A.}\ \bibnamefont {Oesch}}, \bibinfo {author} {\bibfnamefont {A.}~\bibnamefont {Weibel}}, \bibinfo {author} {\bibfnamefont {S.}~\bibnamefont {Wuyts}}, \bibinfo {author} {\bibfnamefont {M.}~\bibnamefont {Xiao}}, \bibinfo {author} {\bibfnamefont {R.}~\bibnamefont
  {Bordoloi}}, \bibinfo {author} {\bibfnamefont {R.}~\bibnamefont {Bouwens}}, \bibinfo {author} {\bibfnamefont {P.}~\bibnamefont {{van Dokkum}}}, \bibinfo {author} {\bibfnamefont {G.}~\bibnamefont {Illingworth}}, \bibinfo {author} {\bibfnamefont {I.}~\bibnamefont {Kramarenko}}, \bibinfo {author} {\bibfnamefont {M.~V.}\ \bibnamefont {Maseda}}, \bibinfo {author} {\bibfnamefont {C.}~\bibnamefont {Mason}}, \bibinfo {author} {\bibfnamefont {R.~A.}\ \bibnamefont {Meyer}}, \bibinfo {author} {\bibfnamefont {E.~J.}\ \bibnamefont {Nelson}}, \bibinfo {author} {\bibfnamefont {N.~A.}\ \bibnamefont {Reddy}}, \bibinfo {author} {\bibfnamefont {I.}~\bibnamefont {Shivaei}}, \bibinfo {author} {\bibfnamefont {R.~A.}\ \bibnamefont {Simcoe}},\ and\ \bibinfo {author} {\bibfnamefont {M.}~\bibnamefont {Yue}},\ }\href {https://doi.org/10.3847/1538-4357/ad2345} {\bibfield  {journal} {\bibinfo  {journal} {ApJ}\ }\textbf {\bibinfo {volume} {963}},\ \bibinfo {pages} {129} (\bibinfo {year} {2024})}\BibitemShut {NoStop}%
\bibitem [{\citenamefont {Greene}\ \emph {et~al.}(2024)\citenamefont {Greene}, \citenamefont {Labbe}, \citenamefont {Goulding}, \citenamefont {Furtak}, \citenamefont {Chemerynska}, \citenamefont {Kokorev}, \citenamefont {Dayal}, \citenamefont {Volonteri}, \citenamefont {Williams}, \citenamefont {Wang}, \citenamefont {Setton}, \citenamefont {Burgasser}, \citenamefont {Bezanson}, \citenamefont {Atek}, \citenamefont {Brammer}, \citenamefont {Cutler}, \citenamefont {Feldmann}, \citenamefont {Fujimoto}, \citenamefont {Glazebrook}, \citenamefont {{de Graaff}}, \citenamefont {Khullar}, \citenamefont {Leja}, \citenamefont {Marchesini}, \citenamefont {Maseda}, \citenamefont {Matthee}, \citenamefont {Miller}, \citenamefont {Naidu}, \citenamefont {Nanayakkara}, \citenamefont {Oesch}, \citenamefont {Pan}, \citenamefont {Papovich}, \citenamefont {Price}, \citenamefont {{van Dokkum}}, \citenamefont {Weaver}, \citenamefont {Whitaker},\ and\ \citenamefont {Zitrin}}]{greene_uncover_2024}%
  \BibitemOpen
  \bibfield  {author} {\bibinfo {author} {\bibfnamefont {J.~E.}\ \bibnamefont {Greene}}, \bibinfo {author} {\bibfnamefont {I.}~\bibnamefont {Labbe}}, \bibinfo {author} {\bibfnamefont {A.~D.}\ \bibnamefont {Goulding}}, \bibinfo {author} {\bibfnamefont {L.~J.}\ \bibnamefont {Furtak}}, \bibinfo {author} {\bibfnamefont {I.}~\bibnamefont {Chemerynska}}, \bibinfo {author} {\bibfnamefont {V.}~\bibnamefont {Kokorev}}, \bibinfo {author} {\bibfnamefont {P.}~\bibnamefont {Dayal}}, \bibinfo {author} {\bibfnamefont {M.}~\bibnamefont {Volonteri}}, \bibinfo {author} {\bibfnamefont {C.~C.}\ \bibnamefont {Williams}}, \bibinfo {author} {\bibfnamefont {B.}~\bibnamefont {Wang}}, \bibinfo {author} {\bibfnamefont {D.~J.}\ \bibnamefont {Setton}}, \bibinfo {author} {\bibfnamefont {A.~J.}\ \bibnamefont {Burgasser}}, \bibinfo {author} {\bibfnamefont {R.}~\bibnamefont {Bezanson}}, \bibinfo {author} {\bibfnamefont {H.}~\bibnamefont {Atek}}, \bibinfo {author} {\bibfnamefont {G.}~\bibnamefont {Brammer}}, \bibinfo {author} {\bibfnamefont
  {S.~E.}\ \bibnamefont {Cutler}}, \bibinfo {author} {\bibfnamefont {R.}~\bibnamefont {Feldmann}}, \bibinfo {author} {\bibfnamefont {S.}~\bibnamefont {Fujimoto}}, \bibinfo {author} {\bibfnamefont {K.}~\bibnamefont {Glazebrook}}, \bibinfo {author} {\bibfnamefont {A.}~\bibnamefont {{de Graaff}}}, \bibinfo {author} {\bibfnamefont {G.}~\bibnamefont {Khullar}}, \bibinfo {author} {\bibfnamefont {J.}~\bibnamefont {Leja}}, \bibinfo {author} {\bibfnamefont {D.}~\bibnamefont {Marchesini}}, \bibinfo {author} {\bibfnamefont {M.~V.}\ \bibnamefont {Maseda}}, \bibinfo {author} {\bibfnamefont {J.}~\bibnamefont {Matthee}}, \bibinfo {author} {\bibfnamefont {T.~B.}\ \bibnamefont {Miller}}, \bibinfo {author} {\bibfnamefont {R.~P.}\ \bibnamefont {Naidu}}, \bibinfo {author} {\bibfnamefont {T.}~\bibnamefont {Nanayakkara}}, \bibinfo {author} {\bibfnamefont {P.~A.}\ \bibnamefont {Oesch}}, \bibinfo {author} {\bibfnamefont {R.}~\bibnamefont {Pan}}, \bibinfo {author} {\bibfnamefont {C.}~\bibnamefont {Papovich}}, \bibinfo {author}
  {\bibfnamefont {S.~H.}\ \bibnamefont {Price}}, \bibinfo {author} {\bibfnamefont {P.}~\bibnamefont {{van Dokkum}}}, \bibinfo {author} {\bibfnamefont {J.~R.}\ \bibnamefont {Weaver}}, \bibinfo {author} {\bibfnamefont {K.~E.}\ \bibnamefont {Whitaker}},\ and\ \bibinfo {author} {\bibfnamefont {A.}~\bibnamefont {Zitrin}},\ }\href {https://doi.org/10.3847/1538-4357/ad1e5f} {\bibfield  {journal} {\bibinfo  {journal} {ApJ}\ }\textbf {\bibinfo {volume} {964}},\ \bibinfo {pages} {39} (\bibinfo {year} {2024})}\BibitemShut {NoStop}%
\bibitem [{\citenamefont {Serot}\ and\ \citenamefont {Walecka}(1992)}]{ainsworth_relativistic_1992}%
  \BibitemOpen
  \bibfield  {author} {\bibinfo {author} {\bibfnamefont {B.~D.}\ \bibnamefont {Serot}}\ and\ \bibinfo {author} {\bibfnamefont {J.~D.}\ \bibnamefont {Walecka}},\ }in\ \href {https://doi.org/10.1007/978-1-4615-3466-2_5} {\emph {\bibinfo {booktitle} {Recent {{Progress}} in {{Many-Body Theories}}}}},\ \bibinfo {editor} {edited by\ \bibinfo {editor} {\bibfnamefont {T.~L.}\ \bibnamefont {Ainsworth}}, \bibinfo {editor} {\bibfnamefont {C.~E.}\ \bibnamefont {Campbell}}, \bibinfo {editor} {\bibfnamefont {B.~E.}\ \bibnamefont {Clements}},\ and\ \bibinfo {editor} {\bibfnamefont {E.}~\bibnamefont {Krotscheck}}}\ (\bibinfo  {publisher} {Springer US},\ \bibinfo {year} {1992})\ pp.\ \bibinfo {pages} {49--92}\BibitemShut {NoStop}%
\bibitem [{\citenamefont {Bowers}\ \emph {et~al.}(1975)\citenamefont {Bowers}, \citenamefont {Gleeson},\ and\ \citenamefont {Pedigo}}]{bowers_relativistic_1975}%
  \BibitemOpen
  \bibfield  {author} {\bibinfo {author} {\bibfnamefont {R.~L.}\ \bibnamefont {Bowers}}, \bibinfo {author} {\bibfnamefont {A.~M.}\ \bibnamefont {Gleeson}},\ and\ \bibinfo {author} {\bibfnamefont {R.~D.}\ \bibnamefont {Pedigo}},\ }\href {https://doi.org/10.1103/PhysRevD.12.3043} {\bibfield  {journal} {\bibinfo  {journal} {Phys. Rev. D}\ }\textbf {\bibinfo {volume} {12}},\ \bibinfo {pages} {3043} (\bibinfo {year} {1975})}\BibitemShut {NoStop}%
\bibitem [{\citenamefont {Dom{\`e}nech}\ and\ \citenamefont {Sasaki}(2021)}]{domenech_cosmology_2021}%
  \BibitemOpen
  \bibfield  {author} {\bibinfo {author} {\bibfnamefont {G.}~\bibnamefont {Dom{\`e}nech}}\ and\ \bibinfo {author} {\bibfnamefont {M.}~\bibnamefont {Sasaki}},\ }\href {https://doi.org/10.1088/1475-7516/2021/06/030} {\bibfield  {journal} {\bibinfo  {journal} {J. Cosmol. Astropart. Phys.}\ }\textbf {\bibinfo {volume} {2021}}\bibfield  {number} {\bibinfo  {number} { (06)},\ \bibinfo {pages} {030}},\ }\Eprint {https://arxiv.org/abs/2104.05271} {arXiv:2104.05271 [hep-th]} \BibitemShut {NoStop}%
\bibitem [{\citenamefont {Archidiacono}\ \emph {et~al.}(2022)\citenamefont {Archidiacono}, \citenamefont {Castorina}, \citenamefont {Redigolo},\ and\ \citenamefont {Salvioni}}]{archidiacono_unveiling_2022}%
  \BibitemOpen
  \bibfield  {author} {\bibinfo {author} {\bibfnamefont {M.}~\bibnamefont {Archidiacono}}, \bibinfo {author} {\bibfnamefont {E.}~\bibnamefont {Castorina}}, \bibinfo {author} {\bibfnamefont {D.}~\bibnamefont {Redigolo}},\ and\ \bibinfo {author} {\bibfnamefont {E.}~\bibnamefont {Salvioni}},\ }\href {https://doi.org/10.1088/1475-7516/2022/10/074} {\bibfield  {journal} {\bibinfo  {journal} {J. Cosmol. Astropart. Phys.}\ }\textbf {\bibinfo {volume} {2022}}\bibinfo  {number} { (10)},\ \bibinfo {pages} {074}}\BibitemShut {NoStop}%
\bibitem [{\citenamefont {Savastano}\ \emph {et~al.}(2019)\citenamefont {Savastano}, \citenamefont {Amendola}, \citenamefont {Rubio},\ and\ \citenamefont {Wetterich}}]{savastano_primordial_2019}%
  \BibitemOpen
\bibfield  {number} {  }\bibfield  {author} {\bibinfo {author} {\bibfnamefont {S.}~\bibnamefont {Savastano}}, \bibinfo {author} {\bibfnamefont {L.}~\bibnamefont {Amendola}}, \bibinfo {author} {\bibfnamefont {J.}~\bibnamefont {Rubio}},\ and\ \bibinfo {author} {\bibfnamefont {C.}~\bibnamefont {Wetterich}},\ }\href {https://doi.org/10.1103/PhysRevD.100.083518} {\bibfield  {journal} {\bibinfo  {journal} {Phys. Rev. D}\ }\textbf {\bibinfo {volume} {100}},\ \bibinfo {pages} {083518} (\bibinfo {year} {2019})},\ \Eprint {https://arxiv.org/abs/1906.05300} {arXiv:1906.05300 [astro-ph]} \BibitemShut {NoStop}%
\bibitem [{\citenamefont {Quiros}(1999)}]{quiros_finite_1999}%
  \BibitemOpen
  \bibfield  {author} {\bibinfo {author} {\bibfnamefont {M.}~\bibnamefont {Quiros}},\ }\href {https://doi.org/10.48550/arXiv.hep-ph/9901312} {\bibinfo {title} {Finite temperature field theory and phase transitions}} (\bibinfo {year} {1999}),\ \Eprint {https://arxiv.org/abs/hep-ph/9901312} {arXiv:hep-ph/9901312} \BibitemShut {NoStop}%
\bibitem [{\citenamefont {Kapusta}\ and\ \citenamefont {Gale}(2023)}]{kapusta_finite-temperature_2023}%
  \BibitemOpen
  \bibfield  {author} {\bibinfo {author} {\bibfnamefont {J.~I.}\ \bibnamefont {Kapusta}}\ and\ \bibinfo {author} {\bibfnamefont {C.}~\bibnamefont {Gale}},\ }\href {https://doi.org/10.1017/9781009401968} {\emph {\bibinfo {title} {Finite-{{Temperature Field Theory}}: {{Principles}} and {{Applications}}}}},\ \bibinfo {edition} {2nd}\ ed.\ (\bibinfo  {publisher} {Cambridge University Press},\ \bibinfo {year} {2023})\BibitemShut {NoStop}%
\bibitem [{\citenamefont {Fetter}\ and\ \citenamefont {Walecka}(2003)}]{fetter_quantum_2003}%
  \BibitemOpen
  \bibfield  {author} {\bibinfo {author} {\bibfnamefont {A.}~\bibnamefont {Fetter}}\ and\ \bibinfo {author} {\bibfnamefont {J.~D.}\ \bibnamefont {Walecka}},\ }\href@noop {} {\emph {\bibinfo {title} {Quantum {{Theory}} of {{Many-Particle Systems}}}}}\ (\bibinfo  {publisher} {Dover Publications},\ \bibinfo {year} {2003})\BibitemShut {NoStop}%
\bibitem [{\citenamefont {Leggett}(2006)}]{leggett_quantum_2006}%
  \BibitemOpen
  \bibfield  {author} {\bibinfo {author} {\bibfnamefont {A.~J.}\ \bibnamefont {Leggett}},\ }\href@noop {} {\emph {\bibinfo {title} {Quantum Liquids: {{Bose}} Condensation and {{Cooper}} Pairing in Condensed-Matter Systems}}}\ (\bibinfo  {publisher} {Oxford University Press},\ \bibinfo {year} {2006})\BibitemShut {NoStop}%
\bibitem [{\citenamefont {Grimm}(2007)}]{grimm_ultracold_2007}%
  \BibitemOpen
  \bibfield  {author} {\bibinfo {author} {\bibfnamefont {R.}~\bibnamefont {Grimm}},\ }\href@noop {} {\bibinfo {title} {Ultracold {{Fermi}} gases in the {{BEC-BCS}} crossover: A review from the {{Innsbruck}} perspective}} (\bibinfo {year} {2007}),\ \Eprint {https://arxiv.org/abs/cond-mat/0703091} {arXiv:cond-mat/0703091} \BibitemShut {NoStop}%
\bibitem [{\citenamefont {Pisarski}\ and\ \citenamefont {Rischke}(1999)}]{pisarski_superfluidity_1999}%
  \BibitemOpen
  \bibfield  {author} {\bibinfo {author} {\bibfnamefont {R.~D.}\ \bibnamefont {Pisarski}}\ and\ \bibinfo {author} {\bibfnamefont {D.~H.}\ \bibnamefont {Rischke}},\ }\href {https://doi.org/10.1103/PhysRevD.60.094013} {\bibfield  {journal} {\bibinfo  {journal} {Phys. Rev. D}\ }\textbf {\bibinfo {volume} {60}},\ \bibinfo {pages} {094013} (\bibinfo {year} {1999})},\ \Eprint {https://arxiv.org/abs/nucl-th/9903023} {arXiv:nucl-th/9903023} \BibitemShut {NoStop}%
\end{thebibliography}%

\clearpage
\onecolumngrid
\onecolumnfootnotes

\begin{center}
{\large\bfseries SUPPLEMENTAL MATERIAL}
\end{center}

\setcounter{section}{0}
\setcounter{equation}{0}
\setcounter{figure}{0}
\setcounter{table}{0}

\renewcommand{\thesection}{S\arabic{section}}
\renewcommand{\theequation}{S\arabic{equation}}
\renewcommand{\thefigure}{S\arabic{figure}}
\renewcommand{\thetable}{S\arabic{table}}

\section{Density profiles}
\subsection{Profile equations}
\label{sec:profile_equations}

For completeness, we summarize the relativistic equations for a static,
spherically symmetric configuration of degenerate fermions coupled to a
Yukawa mediator and to gravity~\cite{walecka_theory_1974,ainsworth_relativistic_1992,stephenson_jr_neutrino_2012,smirnov_neutrino_2022}. In the parameter space relevant for the
main text, gravitational effects are negligible; the numerical profiles
shown in Fig.~1 were therefore obtained in the nongravitational limit.

The equations follow from energy-momentum conservation,
$\nabla_\mu T^{\mu}{}_{r}=0$, together with the Einstein equations and the
Klein-Gordon equation for the scalar field. We model the degenerate fermions
as a perfect fluid,
\begin{equation}
    T^{\mu\nu}_\psi
    =
    (\rho_\psi+P_\psi)u^\mu u^\nu
    -
    P_\psi g^{\mu\nu},
    \label{eq:T_psi}
\end{equation}
with energy density and pressure given by degenerate integrals
\begin{align}
    \rho_\psi
    &=
    \frac{1}{\pi^2}
    \int_0^{p_F}
    p^2\,dp\,\sqrt{p^2+m_{\psi,\rm eff}^2},
    \label{eq:rho}
    \\
    P_\psi
    &=
    \frac{1}{3\pi^2}
    \int_0^{p_F}
    p^2\,dp\,
    \frac{p^2}{\sqrt{p^2+m_{\psi,\rm eff}^2}} .
    \label{eq:P}
\end{align}
Here we assume Dirac fermions with $g_\psi=2$ and
%Here
\begin{equation}
    m_{\psi,\rm eff} \equiv m_\psi + g\phi ,
\end{equation}
so that the Yukawa interaction is included through the effective
fermion mass.

We use the static, spherically symmetric metric convention
\begin{equation}
    ds^2
    =
    e^{2\Phi(r)}dt^2
    -
    e^{2\Lambda(r)}dr^2
    -
    r^2d\Omega^2,
    \qquad
    e^{-2\Lambda(r)}
    \equiv
    1-\frac{2G M(r)}{r},
    \label{eq:metric}
\end{equation}
and $u^\mu=(e^{-\Phi},0,0,0)$. The diagonal components of the Einstein equation, $G^t{}_t=8\pi G\,T^t{}_t$ and $G^r{}_r=8\pi G\,T^r{}_r$, give
\begin{equation}
    M'(r)
    =
    4\pi r^2 \rho_{\rm tot}(r),
    \qquad
    \Phi'(r)
    =
    \frac{
    G M(r)+4\pi G r^3 p_{r,\rm tot}(r)
    }{
    r\,[r-2G M(r)]
    } .
    \label{eq:einstein}
\end{equation}
The total energy density and radial pressure are
\begin{equation}
    \rho_{\rm tot}
    =
    \rho_\psi+\rho_\phi,
    \qquad
    p_{r,\rm tot}
    =
    P_\psi+p_{r,\phi},
\end{equation}
with the fermion contribution given by the above integral, and for the scalar field
\begin{equation}
    \rho_\phi
    =
    \frac12
    \left(
    e^{-2\Lambda}\phi'^2
    +
    m_\phi^2\phi^2
    \right),
    \qquad
    p_{r,\phi}
    =
    \frac12
    \left(
    e^{-2\Lambda}\phi'^2
    -
    m_\phi^2\phi^2
    \right).
    \label{eq:scalar_stress}
\end{equation}
The scalar Klein-Gordon equation
\begin{equation}
    \phi''
    +
    \left(
    \Phi'-\Lambda'+\frac{2}{r}
    \right)\phi'
    -
    e^{2\Lambda}m_\phi^2\phi
    =
    e^{2\Lambda}g\,\tilde n_\psi ,
    \label{eq:KG_curved}
\end{equation}
where the scalar density is
\begin{equation}
    \tilde n_\psi
    \equiv
    \langle \bar\psi\psi\rangle
    =
    \frac{1}{\pi^2}
    \int_0^{p_F}
    p^2\,dp\,
    \frac{m_{\psi,\rm eff}}
    {\sqrt{p^2+m_{\psi,\rm eff}^2}} .
    \label{eq:ntilde}
\end{equation}
Finally, the radial component of energy-momentum conservation gives
\begin{equation}
    P_\psi'(r)
    =
    -
    \left(\rho_\psi+P_\psi\right)\Phi'(r)
    -
    g\,\tilde n_\psi\,\phi'(r).
    \label{eq:P_continuity}
\end{equation}
In the non-relativistic limit, $\tilde n_\psi\simeq n_\psi$,
$\rho_\psi\simeq m_\psi n_\psi$, and $P_\psi\ll\rho_\psi$, which yields
Eq.~(2) of the main text.

\subsection{Solution method and boundary conditions}
\label{sec:solution_method}

Equations~\eqref{eq:einstein}, \eqref{eq:KG_curved}, and
\eqref{eq:P_continuity} define the radial profile. Since the dwarf-galaxy
solutions studied here are dominated by the Yukawa attraction, we solve the
profiles in the limit $\Phi=\Lambda=0$. In this limit the equations reduce to
\begin{align}
    M'(r)
    &=
    4\pi r^2\rho_{\rm tot}(r),
    \label{eq:M_flat}
    \\
    \phi''
    +
    \frac{2}{r}\phi'&
    -
    m_\phi^2\phi
    =
    g\,\tilde n_\psi,
    \label{eq:KG_flat}
    \\
    P_\psi'(r)
    &=
    -g\,\tilde n_\psi\,\phi'(r).
    \label{eq:P_flat}
\end{align}
Solving the full general-relativistic system gives indistinguishable profiles in the parameter range shown in the main text. The last equation can equivalently be obtained from the requirement of chemical equilibrium $\frac{d\mu}{dr}=0$, as done in previous work~\cite{smirnov_neutrino_2022}.

The unknown functions may be taken to be $M(r)$, $\phi(r)$, and $p_F(r)$.
Because the scalar equation is second order, four boundary conditions are
required. The outer radius $R$ of the configuration is not fixed in advance;
it is determined by the condition that the fermion density vanishes at the
surface.
For $r>R$, the fermion density is zero and the scalar field obeys the
homogeneous massive Klein-Gordon equation. Imposing $\phi(r\to\infty)=0$
gives the exterior solution
\begin{equation}
    \phi(r\ge R)
    =
    \phi(R)\,
    \frac{R}{r}
    e^{-m_\phi(r-R)} .
    \label{eq:phi_exterior}
\end{equation}
Taking the derivative at $r=R$ yields\footnote{Ref.~\cite{smirnov_neutrino_2022} imposes the asymptotic condition through an equivalent Green-function integral relation; see their Eq.~(G.6). Here we instead impose it directly at the surface by matching to the unique source-free exterior solution that decays at infinity, which gives Eq.~\eqref{eq:phi_R}.}
\begin{equation}
    \phi'(R)
    =
    -
    \left(
    m_\phi+\frac{1}{R}
    \right)
    \phi(R).
    \label{eq:phi_R}
\end{equation}
Together with regularity at the origin,
\begin{equation}
    M(0)=0,
    \qquad
    \phi'(0)=0,
\end{equation}
and the fixed total mass condition,
\begin{equation}
    M(R)=M,
\end{equation}
this defines a free-boundary  problem on $0\le r\le R$.

We solve this system numerically using two complementary methods. In most of
the parameter space, a shooting method is sufficient. One chooses trial values
for $p_F(0)$ and $\phi(0)$, integrates outward from the origin, and defines
$R$ by the point where $p_F$ reaches zero. The trial values are then adjusted
until the exterior matching condition~\eqref{eq:phi_R} and the desired total
mass are satisfied.

At large $M$, where the fermions become relativistic, the
shooting method becomes numerically unstable (the reason will be explained below). In this regime, we solve the
same boundary-value problem using a standard collocation-based
boundary-value solver. Such solvers replace the differential equations by a
set of algebraic equations imposed on a finite mesh, and solve these
equations simultaneously with the boundary conditions. We treat the unknown
radius $R$ as an additional parameter, fixed by the surface condition
$p_F(R)=0$. The collocation solver requires a good initial guess to converge. We therefore initialized  it with a stable non-relativistic shooting
solution and then varied the parameters in small steps, using each converged
profile as the initial guess for the next one.
The shooting method is thus still required to seed the collocation solver and, in the nonrelativistic regime, is generally more robust. Whenever both methods converge, we have verified that they yield the same solution.

To obtain the Tremaine--Gunn bound shown in Fig.~\ref{fig:parspace} of the main text, we determined the model parameters for which an equilibrium configuration with $M=10^8\,M_\odot$ and $R=1\,{\rm kpc}$ exists. For the left panel, we scanned over the coupling strength $\alpha$ and, at each value, varied $m_\psi$, repeatedly solving for a profile with fixed mass $M=10^8\,M_\odot$, until the solution had $R=1\,{\rm kpc}$. For the right panel, we analogously scanned over $m_\phi$ and adjusted $\alpha$. When $m_\phi^{-1}\ll 1\,{\rm kpc}$, however, equilibrium solutions with $R=1\,{\rm kpc}$ cease to exist~\cite{smirnov_neutrino_2022}. This regime is indicated by the black vertical line in the figure. Beyond this point, we therefore decreased the target radius from $1\,{\rm kpc}$ until a converged $M=10^8\,M_\odot$ solution could be obtained. The $R=1\,{\rm kpc}$ criterion can no longer be imposed there because no such equilibrium solution exists; we therefore define the continuation of the bound using the largest radius for which a converged $M=10^8\,M_\odot$ solution can be found.

\subsection{Large-mass relativistic profiles}
\label{sec:large_mass_profiles}

\begin{figure}
    \centering
    \includegraphics[width=0.8\columnwidth]{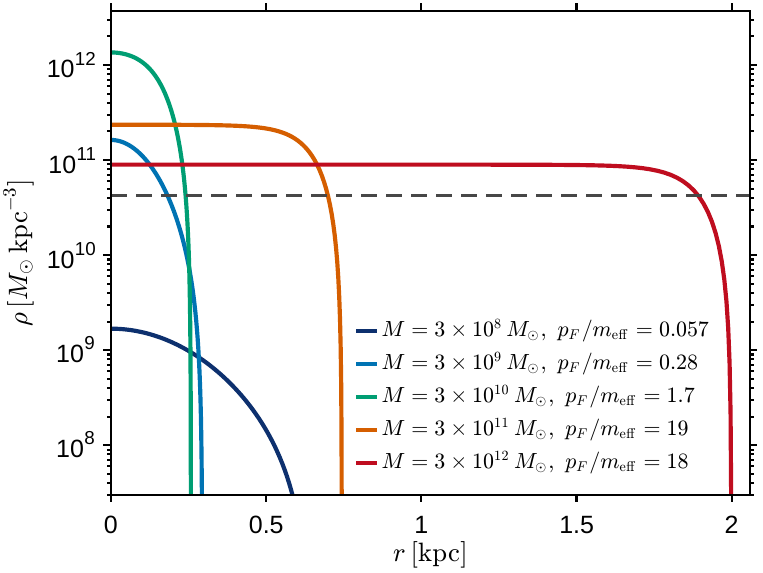}
    \caption{
    Density profiles for several total masses, with
    $m_\phi^{-1}=1\,{\rm kpc}$, $\alpha=3.7\times10^{-51}$, and
    $m_\psi=3\,{\rm eV}$. The legend gives the central value
    $p_F(0)/m_{\psi,\rm eff}(0)$. Profiles with
    $p_F(0)/m_{\psi,\rm eff}(0)\gtrsim 1$ have relativistic fermions near
    the center and approach a nearly constant bulk density. The dashed line shows the analytic large-$M$ limiting bulk density given by Eq.~\eqref{eq:relativistic_bulk_density}.
    }
    \label{fig:profiles_relativistic}
\end{figure}

Figure~\ref{fig:profiles_relativistic} shows density profiles at fixed
$\alpha=3.7\times 10^{-51}$, $m_\phi^{-1}=1\,{\rm kpc}$, and
$m_\psi=3\,{\rm eV}$, while varying the total mass. We characterize the
center of a profile as non-relativistic when
$p_F(0)/m_{\psi,\rm eff}(0)<1$. In this regime, increasing the total mass
raises the central density and decreases the profile radius.
Here $M$ denotes
the total energy of the configuration, including the fermion kinetic energy
and the scalar-field contribution, and is therefore not in general equal to
$N m_\psi$.
Once the central fermions become relativistic, $p_F(0)/m_{\psi,\rm eff}(0)>1$, the
central density no longer grows with $M$. Instead, the profile approaches a
nearly constant bulk density, and additional mass is accommodated mainly by
increasing the radius of the configuration.

The approach to a constant density also explains why the shooting method
becomes unstable in the relativistic regime. The large-$M$ solutions contain
an extended, nearly uniform interior and a comparatively narrow surface
region where the field must match onto the exterior Yukawa tail. As a result,
small changes in the central shooting parameters can lead to large changes in
the surface matching condition. The collocation method avoids this sensitivity
by imposing the differential equations and boundary conditions simultaneously.

\begin{figure}
    \centering
    \includegraphics[width=0.75\columnwidth]{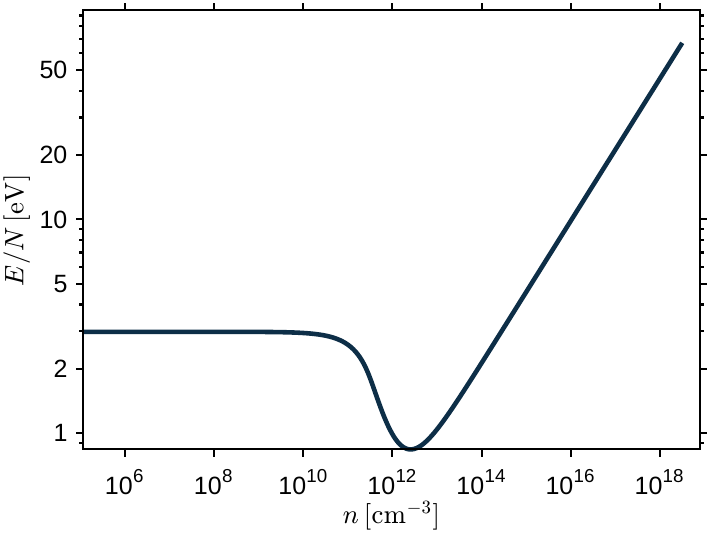}
    \caption{
    Uniform-matter energy per particle as a function of number density for
    the same model parameters as in Fig.~\ref{fig:profiles_relativistic}.
    The minimum occurs at $E/N=0.84\,{\rm eV}$ and
    $n=2.6\times10^{12}\,{\rm cm}^{-3}$, corresponding to an energy density
    $5.7\times10^{10}\,M_\odot\,{\rm kpc}^{-3}$. This minimum sets the
    asymptotic bulk density approached by the large-$M$ profiles.
    }
    \label{fig:binding_energy}
\end{figure}

The limiting density can be understood from the uniform-matter approximation~\cite{walecka_theory_1974}.
For an infinite, constant-density configuration, gradient terms in
Eq.~\eqref{eq:KG_flat} can be neglected, giving
\begin{equation}
\phi
=
-\frac{g}{m_\phi^2}\tilde n_\psi .
\label{eq:phi_uniform}
\end{equation}
For a fixed number density $n$, this equation determines the scalar field
self-consistently, since $\tilde n_\psi$ depends on
$m_{\psi,\rm eff}$. The corresponding total energy per particle,
\begin{equation}
\frac{E}{N}
=
\frac{1}{n}
\left[
\rho_\psi\bigl(n,\phi(n)\bigr)
+
\frac12 m_\phi^2\phi(n)^2
\right],
\label{eq:uniform_energy_per_particle}
\end{equation}
is shown in Fig.~\ref{fig:binding_energy}. Its minimum defines the preferred
bulk density approached by sufficiently massive profiles. 
We stress this uniform-matter estimate does not include gradient or surface effects, and therefore becomes accurate only in the large-$M$ limit.

In the deeply relativistic branch, the preferred bulk density can be
obtained analytically~\cite{walecka_theory_1974,bowers_relativistic_1975,stephenson_jr_neutrino_2012}. For $p_F\gg m_{\psi,\rm eff}$, the scalar density is
$\tilde n_\psi\simeq m_{\psi,\rm eff}p_F^2/(2\pi^2)$, and the uniform field
equation gives
\begin{equation}
    m_{\psi,\rm eff}
    \simeq
    \frac{m_\psi}
    {1+g^2p_F^2/(2\pi^2m_\phi^2)} .
\end{equation}
In the relativistic strong-mean-field regime, defined by 
$g^2p_F^2/(2\pi^2m_\phi^2)\gg1$, we have
$m_{\psi,\rm eff}\ll m_\psi$ and hence
$|g\phi|\simeq m_\psi$. The scalar energy density then approaches
$m_\phi^2\phi^2/2\simeq m_\phi^2m_\psi^2/(2g^2)$, while for a uniform field $P_\phi = -\rho_\phi$. Requiring the total pressure of
the uniform phase to vanish, equivalently minimizing its energy per
particle, gives
\begin{equation}
    \frac{p_F^4}{12\pi^2}
    \simeq
    \frac{m_\phi^2m_\psi^2}{2g^2},
    \qquad
    p_{F,\rm bulk}
    \simeq
    \left(\frac{3\pi}{2\alpha}\right)^{1/4}
    \sqrt{m_\psi m_\phi},
\end{equation}
where $\alpha=g^2/(4\pi)$. The corresponding number density follows from
$n=p_F^3/(3\pi^2)$. Moreover, in the relativistic limit
$P_\psi=\rho_\psi/3$, while for the uniform scalar field
$P_\phi=-\rho_\phi$. The zero-pressure condition therefore implies
$\rho_\psi=3\rho_\phi$, so that
$\rho_{\rm bulk}=\rho_\psi+\rho_\phi=4\rho_\phi$. Hence
\begin{equation}
    n_{\rm bulk}
    \simeq
    \frac{1}{3\pi^2}
    \left(\frac{3\pi}{2\alpha}\right)^{3/4}
    (m_\psi m_\phi)^{3/2},
    \qquad
    \rho_{\rm bulk}
    \simeq
    \frac{m_\psi^2m_\phi^2}{2\pi\alpha}.
    \label{eq:relativistic_bulk_density}
\end{equation}
Thus, neglecting the surface layer and the exterior scalar-field energy,
the large-mass profile approaches the mass-radius relation
\begin{equation}
    R\simeq
    \left(\frac{3M}{4\pi\rho_{\rm bulk}}\right)^{1/3}
    =
    \left(\frac{3\alpha M}
    {2m_\psi^2m_\phi^2}\right)^{1/3}.
\end{equation}
For the benchmark parameters used in
Fig.~\ref{fig:profiles_relativistic}, these expressions give
$p_{F,\rm bulk}\simeq0.83\,{\rm eV}$,
$n_{\rm bulk}\simeq2.5\times10^{12}\,{\rm cm}^{-3}$, and
$\rho_{\rm bulk}\simeq5.4\times10^{10}\,
M_\odot\,{\rm kpc}^{-3}$. The latter is shown as the dashed line in the plot. 

\section{Model building}

In this section we describe in detail a theory which has a long-ranged Yukawa force between DM fermions, $\psi$, that is inactive until late times. The complete interaction Lagrangian we consider is 
\begin{equation}
    \mathcal{L} \supset
    % = \frac{1}{2}(\partial\phi)^2 + \frac{1}{2}(\partial \chi)^2 
    - \frac{1}{2}m_\phi^2 \phi^2
    -\frac{1}{2}m_\chi^2 \chi^2 
    - \frac{1}{4}\lambda\chi^4
    - \frac{\kappa}{2}\phi^2\chi^2
    - g\phi\bar{\psi}\psi 
    + \frac{1}{2F}\chi^2\bar{\psi}\psi .
\end{equation}
This description is valid up to $\Lambda_\text{UV}\sim \text{keV}$, chosen to exceed the largest mass in the theory, $m_\psi$. The dimension-5 operator arises from integrating out a heavy field which mixes the $\chi$ and $\psi$ sectors through weak couplings, with an effective suppression scale $F\sim \text{TeV} \gg \Lambda_\text{UV}$. We shall see below that smaller values for $F$ are viable, but worsen the fine-tuning.
As described in the main text, in order to form dwarf galaxies out of fermionic constituents of mass $m_\psi\sim 1\text{ eV}$, we must fix $g\sim 10^{-25}$ and $m_\phi\sim 10^{-27}\text{ eV}$ to ensure that the force is kpc-ranged and a few decades stronger than gravity. The remaining parameters are determined by the trigger mechanism. 

\subsection{Density-triggered evolution}

We can track the cosmological evolution of the scalars by considering their potential in the presence of a homogeneous cosmological density $\left<\bar{\psi}\psi\right>\approx n_\psi$\footnote{
The approximation is only valid when the fermions are non-relativistic, which is true at least during matter domination, the time frame of interest here.
Regardless, the relativistic density $\langle\bar{\psi}\psi\rangle \sim m_\psi p_F^2 \propto z^2$ still grows monotonically, ensuring that $\chi$ remains in its broken phase and keeping the $\phi$ force turned off, even at earlier times. If production happens sufficiently early, the density can even be large enough to cancel the bare $m_\psi$ entirely.
}:
\begin{equation}
    V(\chi,\phi) = \frac{1}{2}m_\chi^2\chi^2 + \frac{1}{4}\lambda\chi^4
    - \frac{n_\psi}{2F}\chi^2
    + \frac{1}{2}m_\phi^2\phi^2
    + gn_\psi\phi
    + \frac{1}{2}\kappa\chi^2\phi^2.
\end{equation}
From this we can read off the effective mass for $\chi$:
\begin{equation}\label{eq:mchi-eff}
    m_{\chi,\text{eff}}^2 = m_\chi^2  + \kappa\phi^2- \frac{n_\psi}{F}.
\end{equation}
At early times, the last of these terms will dominate, giving the potential a nontrivial minimum
\begin{align}
    \chi_*^2 &= \frac{-m_{\chi,\text{eff}}^2}{\lambda} 
    \sim \frac{n_\psi}{\lambda F} 
    % &\sim (10\text{ eV})^2\left(\frac{z}{z_\text{eq}}\right)^3\left(\frac{m_\psi}{\text{eV}}\right)^{-1}\left(\frac{\lambda}{10^{-5}}\right)^{-1}\left(\frac{F}{\text{keV}}\right)^{-1} \\
    \sim (10^6\text{ eV})^2
    \left(\frac{z}{z_\text{eq}}\right)^3
    \left(\frac{\lambda}{10^{-25}}\right)^{-1}
    \left(\frac{F}{\text{TeV}}\right)^{-1}
    \left(\frac{m_\psi}{\text{eV}}\right)^{-1}.
\end{align}
The auxiliary scalar thus acquires a vacuum expectation value, as it is heavy enough to track this minimum, $m_{\chi,\text{eff}}\gg H$. Integrating out this field, the potential for the Yukawa scalar can be computed as
\begin{align}
    U(\phi)=V(\chi_*,\phi) &= gn_\psi \phi + \frac{1}{2}m_\phi^2\phi^2  + \frac{1}{2}m_\chi^2\chi_*^2 + \frac{1}{4}\lambda\chi_*^4
    + \frac{1}{2}\kappa\chi_*^2\phi^2 - \frac{1}{2}\chi_*^2\left( m_\chi^2 + \kappa\phi^2 + \lambda\chi_*^2 \right) \\
    &= gn_\psi \phi + \frac{1}{2}m_\phi^2\phi^2 - \frac{1}{4}\lambda\chi_*^4\\
    &=  gn_\psi\,\phi + \frac{1}{2}\left(m_\phi^2 + \frac{\kappa}{\lambda}\left(\frac{n_\psi}{F} - m_\chi^2\right)\right)\phi^2
    - \frac{\kappa^2}{4\lambda}\phi^4 + \text{const.} \label{eq:U-phi}
\end{align}
(Note that to simplify, in the first line we have made the substitution $n_\psi/F = m_\chi^2 + \kappa \phi^2 + \lambda \chi_*^2$ based on the definition of $\chi_*$ above.) This means that the ultralight scalar experiences finite-density corrections to its mass at early times:
\begin{align}
    m_{\phi,\text{eff}}^2 &= m_\phi^2 + \frac{\kappa}{\lambda}\left(\frac{n_\psi}{F} - m_\chi^2\right) 
    \sim \frac{\kappa \, n_\psi}{\lambda F} \nonumber\\
    &\sim (10^{-6}\text{ eV})^2
    \left(\frac{z}{z_\text{eq}}\right)^3
    \left(\frac{\lambda}{10^{-25}}\right)^{-1}
    \left(\frac{\kappa}{10^{-24}}\right)
    \left(\frac{F}{\text{TeV}}\right)^{-1}
    \left(\frac{m_\psi}{\text{eV}}\right)^{-1}.
\end{align}
The $\phi$-mediated force is inactive since the corresponding Jeans length $\ell_{J,\phi}$ exceeds its range:
\begin{align}
    1 \lesssim \frac{\ell_{J,\phi}^2}{m_{\phi,\text{eff}}^{-2}} 
    &\sim \frac{m_\psi^2v_F^2}{g^2 m_\psi n_\psi} \frac{\kappa n_\psi}{\lambda F}
    \sim \frac{\kappa\,n_\psi^{2/3}(z_\text{form})}{g^2 \lambda m_\psi F} \nonumber \\
    &\sim 10^{35}
    \left(\frac{z_\text{form}}{100}\right)^2
    \left(\frac{\lambda}{10^{-25}}\right)^{-1}
    \left(\frac{\kappa}{10^{-24}}\right)
    \left(\frac{F}{\text{TeV}}\right)^{-1}
    \left(\frac{g}{10^{-25}}\right)^{-2}
    \left(\frac{m_\psi}{\text{eV}}\right)^{-5/3}.
    % \iff \frac{\kappa}{\lambda} \gtrsim \frac{g^2 m_\psi F}{n_\psi^{2/3}(z_\text{form})}
    % \sim 10^{-32}
    % \left(\frac{z_\text{form}}{100}\right)^{-2}
    % \left(\frac{F}{\text{PeV}}\right)
    % \left(\frac{g}{10^{-25}}\right)^{2}
    % \left(\frac{m_\psi}{\text{eV}}\right)^{5/3}.
\end{align}

The bare mass of the auxiliary field is chosen such that the effective mediator mass reverts to being ultralight precisely at the desired redshift $z_\text{form}$:
\begin{equation}
     m_\chi^2 = \frac{n_\psi(z_\text{form})}{F} 
     % \sim (10^{-4}\text{ eV})^2
     % \left(\frac{z_\text{form}}{100}\right)^3
     % \left(\frac{m_\psi}{\text{eV}}\right)^{-1}
     % \left(\frac{F}{\text{keV}}\right)^{-1}
     \sim (10^{-9}\text{ eV})^2
     \left(\frac{z_\text{form}}{100}\right)^3
     \left(\frac{F}{\text{TeV}}\right)^{-1}
     \left(\frac{m_\psi}{\text{eV}}\right)^{-1}.
\end{equation}
 The transition $m_{\phi\text{,eff}}\to m_\phi$ happens smoothly, since $\chi$ tracks its minimum adiabatically as it moves to zero.
Subsequently, the Yukawa force is effectively turned on, and the above picture breaks down due to rapid bound state formation. Since the force is triggered by finite density, it must shut off above $n_\psi(z_\text{form})$ by construction. At late times, collapsing regions therefore cannot compress beyond that limit (except gravitationally), which self-regulates the maximum density of bound states. A corollary is that the scalar vev contributes a negligible fraction of the DM density in \emph{any} DM-dominated structure (as this is a weaker condition than Eq.~\eqref{eq:phi-edensity}, below).

It is important to note that $U(\phi)$ has its regime of validity set by the maximum field excursion $\phi_c^2 \sim n_\psi/\kappa F$ beyond which $\chi$ returns to its symmetric phase, based on Eq.~\eqref{eq:mchi-eff}. This turns out to parametrically be the same as the classical turning point $\phi_\text{tp}^2 \sim m_{\phi,\text{eff}}^2(\kappa^2/\lambda)^{-1}$, beyond which the potential turns over due to the negative quartic term. There is thus a stable minimum at $\phi_*\sim gn_\psi/m_{\phi,\text{eff}}^2$ provided that
\begin{align}
    1 \gtrsim \frac{\phi_*^2}{\phi_\text{tp}^2} \sim\frac{g^2 n_\psi^2\left(\frac{\kappa \, n_\psi}{\lambda F} \right)^{-2}}{\frac{\lambda}{\kappa^2} \left(\frac{\kappa \, n_\psi}{\lambda F} \right)}
    &\gtrsim \frac{g^2\lambda^2 F^3}{\kappa \, n_\psi(z_\text{form})}\nonumber \\
    &\sim  10^{-33}
    \left(\frac{z_\text{form}}{100}\right)^{-3}
    \left(\frac{\lambda}{10^{-25}}\right)^{2}
    \left(\frac{10^{-24}}{\kappa}\right)
    \left(\frac{F}{\text{TeV}}\right)^{3}
    \left(\frac{g}{10^{-25}}\right)^{2}
    \left(\frac{m_\psi}{\text{eV}}\right),
\end{align}
which is easily satisfied. Like the auxiliary scalar, $m_{\phi,\text{eff}}\gg H$, so $\phi$ follows its minimum adiabatically. No back-reaction arises on $\chi$, through the same condition:
\begin{equation}
    \frac{\kappa \phi_*^2}{m_\chi^2}
    \sim \frac{\kappa g^2 n_\psi^2}{\left(\frac{n_\psi}{F}\right) \left(\frac{\kappa\, n_\psi}{\lambda F}\right)^2}
    \lesssim \frac{g^2 \lambda^2 F^3}{\kappa\, n_\psi(z_\text{form})} \ll 1.
\end{equation}
The contribution to the energy budget of the Universe from $\phi$ is negligible, since
\begin{align}
    1&\gtrsim \frac{|U(\phi_*)|}{\rho_\text{DM}} 
    \sim \frac{m_{\phi,\text{eff}}^2\,\phi_*^2}{m_\psi n_\psi }
    \sim \frac{g^2 \lambda F}{\kappa \,m_\psi}\label{eq:phi-edensity}
    \sim 10^{-39}
    \left(\frac{\lambda}{10^{-25}}\right)
    \left(\frac{10^{-24}}{\kappa}\right)
    \left(\frac{F}{\text{TeV}}\right)
    \left(\frac{g}{10^{-25}}\right)^{2}
    \left(\frac{\text{eV}}{m_\psi}\right),
\end{align}
whereas (to ensure early clustering, as we shall see below) the $\chi$-induced binding energy contributes
\begin{equation}
    \frac{|V(\chi_*,\phi)|}{\rho_\text{DM}} 
    \sim \frac{\lambda \chi_*^4}{m_\psi n_\psi} 
    \sim \frac{n_\psi}{\lambda m_\psi  F^2}
    \sim 1
    \left(\frac{z}{z_\text{eq}}\right)^3
    \left(\frac{\lambda}{10^{-25}}\right)^{-1}
    \left(\frac{F}{\text{TeV}}\right)^{-2}
    \left(\frac{m_\psi}{\text{eV}}\right)^{-2}.
\end{equation}
Finally, the scalar expectation values do not affect the fermion mass, since the relevant conditions $g\phi_*,\chi_*^2/F \lesssim m_\psi$ are equivalent to the two inequalities above.

\subsection{Early structures}

Having studied the homogeneous evolution of the scalar sector, we now consider its phenomenology at the perturbative level. In particular, although $\phi$ is reduced to a (negligibly weak) contact interaction at early times, the higher-dimension coupling between $\chi$ and the DM is enhanced in the presence of the coherent condensate $\chi_*\neq 0$, into an effective Yukawa with (time-dependent)  strength
\begin{equation}
    y = \frac{\chi_*}{F} 
    \sim \sqrt{\frac{n_\psi}{\lambda F^3}} 
    \sim 10^{-6}
    \left(\frac{z}{z_\text{eq}}\right)^{3/2}
    \left(\frac{\lambda}{10^{-25}}\right)^{-1/2}
    \left(\frac{F}{\text{TeV}}\right)^{-3/2}
    \left(\frac{m_\psi}{\text{eV}}\right)^{-1/2},
\end{equation}
and whose range can be computed from the curvature of the field around its minimum:
\begin{align}
    m_{\chi,\text{phys}}^2 &= \partial_\chi^2V|_{\chi_*} 
    = 2\lambda \chi_*^2 
    \sim \frac{n_\psi}{F} 
    \sim (10^{-6}\text{ eV})^2
    \left(\frac{z}{z_\text{eq}}\right)^{3}
    \left(\frac{F}{\text{TeV}}\right)^{-1}
    \left(\frac{m_\psi}{\text{eV}}\right)^{-1}, \\
    m_{\chi,\text{phys}}^{-1} &\sim 10\text{ cm}
    \left(\frac{z}{z_\text{eq}}\right)^{-3/2}
    \left(\frac{F}{\text{TeV}}\right)^{1/2}
    \left(\frac{m_\psi}{\text{eV}}\right)^{1/2}.
\end{align}

Linear perturbations obey $\ddot{\delta}_k + 2H\dot{\delta}_k + \omega^2 \delta_k = 0$, with dispersion relation~\cite{nusser_structure_2005,domenech_cosmology_2021,domenech_halo_2023,archidiacono_unveiling_2022}
\begin{equation}
    \omega^2(k)=c_s^2\left(\frac{k}{a}\right)^2 - 4\pi G \bar{\rho} 
    - \frac{y^2 \bar{\rho}}{m_\psi^2}\frac{(k/a)^2}{(k/a)^2+m_{\chi,\text{phys}}^2} 
    - \frac{g^2 \bar{\rho}}{m_\psi^2}\frac{(k/a)^2}{(k/a)^2+m_{\phi,\text{eff}}^2}.
\end{equation}
The last term (corresponding to $\phi$ exchange) can be discarded as it is negligible compared to the $\chi$-mediated force for all $k$, since $g \ll y,m_{\phi,\text{eff}}^{-1} \lesssim m_{\chi,\text{phys}}^{-1}$. 
With only gravitational attraction, the $\psi$ fluid would be a collisionless degenerate gas, and modes shorter than the free-streaming length,
\begin{align}
    \ell_{\text{fs}}^2 &\sim \frac{v_F^2}{H^2} 
    \sim \frac{1}{G m_\psi^3 n_\psi^{1/3}} 
    \sim (80\text{ kpc})^2
    \left(\frac{z}{z_\text{eq}}\right)^{-1}
    \left(\frac{m_\psi}{\text{eV}}\right)^{-8/3},
\end{align}
would be erased~\cite{carena_cosmologically_2022,lesgourgues_massive_2006}, as identified in \textit{(ii)} of the main text. Parametrically, during matter domination $\ell_\text{fs}^2 \sim \ell_{J,\text{grav}}^2 \approx c_s^2/(4\pi G\rho)$, the Jeans length, since $c_s \sim \bar{v}_F, H^2 \sim 4\pi G\bar{\rho}$. 

This undesirable behavior at small scales can be ameliorated in the presence of the early Yukawa, as long as the corresponding Jeans length $\ell_{J,\chi}$ is within the force range, allowing for the rapid formation of nonlinear bound states of size $m_{\chi,\text{phys}}^{-1}$~\cite{savastano_primordial_2019,domenech_halo_2023,domenech_cosmology_2021}, far below astrophysical scales:
\begin{equation}
    1 \gtrsim \frac{\ell_{J,\chi}^2}{m_{\chi,\text{phys}}^{-2}} 
    \sim \frac{m_\psi^2v_F^2}{y^2 m_\psi n_\psi} \frac{n_\psi}{F}
    \sim \frac{\lambda F^2}{m_\psi \,n_\psi^{1/3}(z_\text{form})}
    \sim 1
    \left(\frac{z_\text{form}}{100}\right)^{-1}
    \left(\frac{\lambda}{10^{-25}}\right)
    \left(\frac{F}{\text{TeV}}\right)^2
    \left(\frac{m_\psi}{\text{eV}}\right)^{-2/3}.
\end{equation}
We leave the detailed phenomenology of these clumps to future work, but note that they must necessarily disintegrate after $z_\text{form}$, so that the unscreened $\phi$ exchange force takes over and dwarf galaxy sized objects can form. 
The presence of these DM degrees of freedom is undetectable, except insofar as the properties of the fluid are modified under
\begin{align}
    m_\psi \to M_{\chi\text{bd}} 
    \approx \frac{4}{3} \pi \bar{\rho}\,(m_{\chi,\text{phys}}^{-1})^3 
    \sim \frac{m_\psi^{3/2} F^{3/2}}{\bar{\rho}^{1/2}} 
    &\sim 10^{-15}\text{ g}\,
    \left(\frac{z}{z_\text{eq}}\right)^{-3/2}
    \left(\frac{F}{\text{TeV}}\right)^{3/2}
    \left(\frac{m_\psi}{\text{eV}}\right)^{3/2} ,\\
    n_\psi \to \frac{\bar{\rho}}{M_{\chi\text{bd}}}
    \sim \frac{\bar{\rho}^{3/2}}{m_\psi^{3/2} F^{3/2}} 
    &\sim 10 \text{ m}^{-3}
    \left(\frac{z}{z_\text{eq}}\right)^{9/2}
    \left(\frac{F}{\text{TeV}}\right)^{-3/2}
    \left(\frac{m_\psi}{\text{eV}}\right)^{-3/2}.
\end{align}
The early clumps are thus observationally inert: their mass $M_{\chi\text{bd}}$ lies over thirty orders of magnitude below the lightest scales probed by microlensing or other compact-object searches, and so remains safe even if these parameters evolve alongside the background until $z_\text{form}$.
At the same time, bundling the fermions into these heavy structures reduces the sound speed to the inter-clump velocity dispersion, which is likely to be far below $\bar{v}_F$ (though the details depend on dissipation dynamics that are not considered here). This ensures that the relevant Jeans length is sufficiently short as to not disrupt small-scale structure growth.

\subsection{Radiative stability}

By inspection the small couplings $g,\kappa$ are technically natural but the mediator mass is not.  The shift symmetry $\phi\to\phi + \text{const.}$ that would forbid it is explicitly broken by both the Yukawa and the portal couplings, inducing a self-energy
\begin{equation}
    \frac{(\delta m_\phi^2)_\text{vac}}{m_\phi^2} 
    \sim \frac{\max(g^2,\kappa)\,\Lambda_\text{UV}^2}{16\pi^2 m_\phi^2} 
    % \max\left(\frac{\kappa \, \Lambda_\text{UV}^2}{16\pi^2 m_\phi^2},\frac{g^2 \Lambda_\text{UV}^2}{16\pi ^2m_\phi^2}\right)
    \sim 10^{34}
    \left(\frac{\kappa}{10^{-24}}\right)
    \left(\frac{\Lambda_\text{UV}}{\text{keV}}\right)^2
    \left(\frac{m_\phi}{10^{-27}\text{ eV}}\right)^{-2}.
\end{equation}
Similarly, the mass of the auxiliary scalar is corrected to
\begin{align}
    \frac{(\delta m_\chi^2)_\text{vac}}{m_\chi^2} 
    &\sim \max\left(\frac{m_\psi \Lambda_\text{UV}^2}{16\pi^2m_\chi^2F},\frac{\kappa\,\Lambda_\text{UV}^2}{16\pi^2 m_\chi^2}\right) 
    \sim \frac{\max(m_\psi,\kappa F)\, \Lambda_\text{UV}^2}{16\pi^2n_\psi(z_\text{form})}
    % \sim \max\left(\frac{m_\psi \Lambda_\text{UV}^2}{16\pi^2n_\psi(z_\text{form})},\frac{\kappa\,\Lambda_\text{UV}^2 F}{16\pi^2 n_\psi(z_\text{form})}\right) 
    \sim 10^9
    % \left(\frac{\text{PeV}}{F}\right)
    \left(\frac{z_\text{form}}{100}\right)^{-3}
    \left(\frac{\Lambda_\text{UV}}{\text{keV}}\right)^2
    % \left(\frac{m_\chi}{10^{-10}\text{ eV}}\right)^{-2}
    \left(\frac{m_\psi}{\text{eV}}\right)^2,
\end{align}
from loops of $\psi,\phi$ respectively (with the latter being negligible). 
In addition, both fields acquire a quartic
\begin{equation}
    (\delta \lambda)_\text{portal} \sim \frac{\kappa^2}{16\pi^2}  
    \sim 10^{-50}
    \left(\frac{\kappa}{10^{-24}}\right)^2.
\end{equation}
This is too feeble to correct the relevant coupling $\sim \kappa^2/\lambda$ in the $\phi$ potential (see Eq.~\eqref{eq:U-phi}). For the auxiliary scalar only, there is a further quartic correction due to the higher dimension operator, which dominates,
\begin{equation}
    (\delta \lambda)_\text{dim-5} \sim \frac{\Lambda_\text{UV}^2}{16\pi^2 F^2}
    \sim 10^{-20}
    \left(\frac{F}{\text{TeV}}\right)^{-2}
    \left(\frac{\Lambda_\text{UV}}{\text{keV}}\right)^2,
\end{equation}
 showing that at the benchmark, even the bare quantity $\lambda$ is not natural.
In this model, we must begrudgingly accept these tunings as the price of eV-scale fermion DM.  Avoiding such tuning may require additional ingredients, which we postpone to a future study.

If the scalars are internally thermalized at $T_\phi,T_\chi$, besides vacuum corrections their masses are also subject to thermal fluctuations\footnote{This assumption is conservative; since the self-couplings are weak, the sector may well be athermal and cold, in which case these contributions are entirely absent.}~\cite{quiros_finite_1999,kapusta_finite-temperature_2023}:
\begin{equation}
    (\delta m_\phi^2)_\text{therm} \sim \kappa \langle\chi^2\rangle_\text{therm},
    \qquad (\delta m_\chi^2)_\text{therm} \sim \max(\kappa \langle\phi^2\rangle_\text{therm}, \lambda \langle\chi^2\rangle_\text{therm}).
\end{equation}
The former must be compared to $m_\phi^2$ at late times and the latter to $m_{\chi,\text{phys}}^2$ at early times, in line with the respective epochs during which each is relevant to the DM fluid evolution.
We therefore additionally require that the hidden sector carry negligible entropy, which can be fixed by the choice of initial conditions. This will also ensure that there is no substantial dark radiation (and hence $\Delta N_\text{eff}$) from this sector. Concretely, we need
\begin{equation}
    \frac{T_\phi}{T_\text{SM}} \lesssim \frac{m_{\chi,\text{phys}}}{\sqrt{\kappa} \, T_\text{SM}}
    \sim \frac{1}{T_\text{SM}}\sqrt{\frac{n_\psi}{\kappa F}}
    \sim 100\,
    \left(\frac{z}{1}\right)^{1/2}
    \left(\frac{\kappa}{10^{-24}}\right)^{-1/2}
    \left(\frac{F}{\text{TeV}}\right)^{-1/2}
    \left(\frac{m_\psi}{\text{eV}}\right)^{-1/2},
\end{equation}
which is easily satisfied, and
\begin{equation}
    \kappa m_\chi^2e^{-m_\chi/T_\chi} \lesssim m_\phi^2 
    \iff \frac{T_\chi}{T_\text{SM}} 
    \lesssim \frac{\sqrt{n_\psi(z_\text{form})/F}}{T_\text{SM}\ln(\kappa n_\psi(z_\text{form})/m_\phi^2F)}
    % \lesssim \frac{m_\chi}{T_\text{SM}\ln(\kappa m_\chi^2/m_\phi^2)}
    \sim 10^{-10}
    \left(\frac{z}{z_\text{eq}}\right)^{-1}
    \left(\frac{F}{\text{TeV}}\right)^{-1/2}
    \left(\frac{m_\psi}{\text{eV}}\right)^{-1/2}.
\end{equation}
The derived upper bounds are self-consistent with the ultralight scalar being relativistic, i.e., $\langle\phi^2\rangle_\text{therm} \sim T_\phi^2$, and the auxiliary scalar being non-relativistic, such that $\langle\chi^2\rangle_\text{therm} \sim m_\chi^2e^{-m_\chi/T_\chi}$, due to Boltzmann suppression. 

\section{Superfluidity}

In principle, attractive interactions between the constituents of a cold Fermi gas will induce a pairing instability in the ground state \emph{regardless of their strength}, due to Cooper's theorem~\cite{fetter_quantum_2003,leggett_quantum_2006}. Here, working entirely at the parametric level and ignoring $\mathcal{O}(1)$ factors, we show that this can be ignored in the context of long-ranged forces between fermionic DM. The regime can be defined by $m_\phi \ll p_F$, which is equivalent to the condition that the range exceed the interparticle spacing of the DM fluid. Further, since $\alpha m_\psi/m_\phi \ll 1$ throughout the allowed parameter space (e.g. at the benchmark, it is $\sim 10^{-24}$), two-body bound states are impossible and we are firmly in the BCS limit~\cite{grimm_ultracold_2007}.
The extent of pairing is governed by the size of the gap~\cite{gorkov_contribution_1961,fetter_quantum_2003,pisarski_superfluidity_1999,kapusta_neutrino_2004},
\begin{equation}
    \Delta \approx E_F \exp\left(-\frac{1}{N(0)|\langle\tilde{V}\rangle|}\right),
\end{equation}
where $N(0)$ is the density of states ($\sim m_\psi p_F$ in the non-relativistic regime), and $\langle\tilde{V}\rangle$ is the angularly-averaged momentum-space matrix element for the scattering, both evaluated at the Fermi surface. 
For a long-ranged Yukawa, $\tilde{V}(q) = -g^2/(q^2+m_\phi^2)$, we can estimate the scattering matrix element as its largest (isotropic) contribution:
\begin{equation}
    \langle\tilde{V}\rangle \approx \frac{1}{2}\int_{-1}^1P_{\ell=0}(\cos\theta) \tilde{V} \,d(\cos\theta)  
    \sim - \frac{\alpha}{p_F^2} \ln\left(\frac{p_F}{m_\phi}\right),
\end{equation}
where the log factor captures the enhancement due to forward scattering in this limit. Quantitatively,
\begin{align}
    \frac{1}{N(0)|\langle\tilde{V}\rangle|} &\sim \frac{p_F}{\alpha m_\psi \ln(p_F/m_\phi)} \sim 10^{44} \left(\frac{m_\psi}{\text{eV}}\right)^{-4/3} \left(\frac{\alpha}{10^{-50}}\right)^{-1} \left(\frac{\ln(p_F/m_\phi)}{50}\right)^{-1}.
\end{align}
To be conservative we have evaluated $p_F$ at its minimum value, based on the homogeneous DM density today, in order to reduce the suppression in $\Delta$. 
Notice that the pairing strength is controlled by $\alpha m_\psi/p_F$, which is effectively the ratio of the velocity within a bound pair (if one existed) to the Fermi velocity. This is many orders of magnitude below the parameter setting two-body binding, $\alpha m_\psi /m_\phi$ \cite{garani_condensed_2022}, and farther still from the coherently-enhanced $\alpha m_\psi^2/m_\phi^2$, which determines whether $N$-body bound states are possible (as discussed in the main text).

The gap is thus utterly negligible, as the large exponent means that the pair coherence length $\xi \sim v_F/\Delta$ far exceeds the size of the configuration.
Moreover, any effective DM temperature $T_\text{DM} \sim m_\psi \sigma_v^2$ (due to its cosmological or virial velocity dispersion $\sigma_v$) places the fluid in the unpaired phase, far above the critical temperature $T_c \sim \Delta$. 
We conclude by observing that even if these hurdles are cleared in the context of a different model, the correction entering the equation of state is modest, scaling with $(\Delta / E_F)^2$~\cite{fetter_quantum_2003}, meaning that within the BCS regime, the macroscopic behavior of the DM fluid cannot be qualitatively modified. 

%TC:endignore
\end{document}